\documentclass[11pt]{article}
\usepackage{amsmath}
\usepackage{amssymb}
\usepackage{mathtools}
\usepackage{amsfonts}
\usepackage[english]{babel}
\pdfoutput=1
\usepackage{jheppub}
\usepackage{physics}
\usepackage{comment}
\usepackage{tikz}
\usepackage{bbold} 
\usetikzlibrary{backgrounds,calc}
\usepackage{pgfplots}
\usepackage{bbold}  
\def\centerarc[#1](#2)(#3:#4:#5)
    { \draw[#1] ($(#2)+({#5*cos(#3)},{#5*sin(#3)})$) arc (#3:#4:#5); } 

\usepackage{simplewick} 

\newtheorem{theorem}{Theorem}

\newcommand{\cF}{\mathcal{F}}

\newcommand{\cI}{\mathcal{I}}

\newcommand{\cL}{\mathcal{L}}

\newcommand{\cO}{\mathcal{O}}

\newcommand{\cV}{\mathcal{V}}
\newcommand{\ZZ}{{\mathbb{Z}}}

\title{Exponentiation of higher-point and higher-genus Virasoro conformal blocks in the semiclassical limit}

\author[a]{Marius Gerbershagen,}
\author[b]{Jakob Hollweck}
\affiliation[a]{Theoretische Natuurkunde, Vrije Universiteit Brussel (VUB) and The International Solvay Institutes, Pleinlaan 2, B-1050 Brussels, Belgium}
\affiliation[b]{Theoretisch-Physikalisches Institut, Friedrich-Schiller-Universität Jena,
Max-Wien-Platz 1, D-07743 Jena, Germany}
\emailAdd{marius.gerbershagen@vub.be}
\emailAdd{jakob.hollweck@uni-jena.de}
\abstract{
A long-standing conjecture claims that Virasoro conformal blocks exponentiate in the semiclassical limit $c \to \infty$ with $h/c$ finite.
However, this has been proven only for four-point blocks on the sphere and one-point blocks on the torus.
Here we extend the proof to general conformal blocks for higher-point functions and higher-genus backgrounds in arbitrary channels. The statement is to be understood at the level of a formal power series. Our proof builds upon a novel extension of the oscillator method for the computation of conformal blocks to cases where three internal lines meet at a vertex.
This extension also gives a new constructive method to compute global conformal blocks in 2d CFTs at general genus.
}

\begin{document}

\maketitle

\section{Introduction}

Conformal blocks constitute the basic building blocks of correlation functions in a conformal field theory (CFT).
They determine the part of the correlator which is fixed by conformal symmetry.
An important corner in the parameter space for CFTs is given by the semiclassical limit $c \to \infty$ with all internal and external conformal weights scaling proportional to $c$, meaning that $\lim_{c \to \infty} h/c$ is finite and non-zero.
It has been conjectured that in this limit the conformal blocks exponentiate \cite{Zamolodchikov1986,Zamolodchikov1987},
\begin{equation}
    \begin{aligned}
        &\cF(c,h_1,h_2,\dots,h_p,h_q,\dots;x_1,x_2,\dots)\\
        &\qquad = \exp\left(-\frac{c}{6}\cI\left(\frac{h_1}{c},\frac{h_2}{c},\dots,\frac{h_p}{c},\frac{h_q}{c},\dots;x_1,x_2,\dots\right) + O(c^0)\right)
    \end{aligned}
    \label{eq:exponentiation}
\end{equation}
where $h_1,h_2,\dots$ and $h_p,h_q,\dots$ are the external and internal conformal weights respectively while $x_1,x_2,\dots$ denote the conformal cross-ratios and moduli.
This exponentiation property is of importance for various applications. In holographic CFTs at large central charge, semiclassical Virasoro conformal blocks play a key role in the computation of heavy-light limits, entanglement observables and thermal properties of heavy states (see e.g.~\cite{Hartman:2013mia,Fitzpatrick:2014vua,Fitzpatrick:2015zha}). They also play an important role in the derivation of recursion relations for conformal blocks \cite{Zamolodchikov1987}.
However, up to now exponentiation has been proven only for the special case of the four-point block on the sphere \cite{Besken:2019jyw} and the one-point block on the torus \cite{Desiraju:2024fmo}.\footnote{Apart from the cases of two- and three-point functions on the sphere and the Virasoro character on the torus for which the property trivially holds.}
In this work, we will extend the proof to conformal blocks with an arbitrary number of operator insertions and on arbitrary Riemann surfaces.

\medskip
More precisely, we will prove the following:
\begin{theorem}
  Let $\vec h_\mathrm{ext},\vec h_\mathrm{int}$ be a collection of conformal weights parameterizing the external and internal operators in a given OPE channel and $\vec x$ be a collection of moduli and conformal cross-ratios.
  The conformal block $\cF(c,\vec h_\mathrm{ext},\vec h_\mathrm{int};\vec x)$ in the given channel is defined as a formal power series in $\vec x$.
  Furthermore, let $\vec g_\mathrm{ext},\vec g_\mathrm{int}$ be a collection of positive semiclassical conformal weights.
  Then the following limit exists,
  \begin{equation}
    \cI(\vec g_\mathrm{ext},\vec g_\mathrm{int};\vec x) := -\lim_{c \to \infty}\frac{6}{c}\log\left(\cF(c,c\, \vec g_\mathrm{ext},c\, \vec g_\mathrm{int};\vec x)\right).
    \label{eq:precise-statement-exponentiation}
  \end{equation}
  This determines the semiclassical conformal block $\cI$ as a formal power series in $\vec x$.
\end{theorem}

Our proof largely follows that of \cite{Besken:2019jyw}.
In particular, as in \cite{Besken:2019jyw} the proof makes use of the oscillator method for computing conformal blocks, which was first developed in \cite{Zamolodchikov1986}, to write the conformal block as an integral over an infinite collection of complex-valued oscillator variables.
The integrand decomposes into oscillator wavefunctions which exponentiate individually, making it possible to evaluate the integral by saddle point approximation and thus showing that the conformal block must also exponentiate \cite{Besken:2019jyw}.
The main technical advancement which allows us to extend the proof to arbitrary conformal blocks is a generalization of the oscillator method to three-point functions of arbitrary descendants.
This generalization is necessary in cases where three internal lines of the diagram corresponding to the conformal block meet at a vertex. The machinery developed here may furthermore be useful for higher-genus modular bootstrap, where efficient control over higher-genus conformal blocks is required \cite{Cho:2017oxl}.

Moreover, we provide the following minor improvement on the proof of \cite{Besken:2019jyw}.
While \cite{Besken:2019jyw} proved that exponentiation holds up to a finite set of isolated points, here we also show that this set is empty.
Thus exponentiation holds for any semiclassical conformal block.

Note that for the purpose of this work, we limit ourselves to dealing with conformal blocks as formal power series without taking into account whether these series converge.
It is generally expected that the series expansion of $\cF$ in cross-ratios and moduli (collectively denoted by $\vec x$) converges in a finite region around $\vec x=0$ with singularities at coincident operator insertion points limiting the radius of convergence of both the OPE and the conformal block \cite{Pappadopulo:2012jk} (see e.g.~\cite{2020arXiv200303802G,Arnaudo:2022ivo,LeFloch:2026xec} for more rigorous statements).
For instance, for the four-point block on the sphere the radius of convergence is expected to be $|x|=1$, see e.g.~\cite{Zamolodchikov1987,Felder:2017rgg}.
One similarly expects the series expansion in $\vec x$ of the semiclassical conformal block $\cI$ to converge for small values of $\vec x$ with the radius of convergence limited by poles as well as zeros of the conformal block due to the logarithm in \eqref{eq:precise-statement-exponentiation}.
Moreover, it is possible that multiple semiclassical saddles exist away from $\vec x=0$.
However, we are not aware of general statements on the analytical structure and convergence of semiclassical conformal blocks beyond some special cases such as the one-point block on the torus and the four-point block on the sphere \cite{Desiraju:2024fmo,Menotti:2025pqf}. 

The extended oscillator formalism also provides a new constructive computation method for global conformal blocks (see e.g.~\cite{Antunes:2023kyz,Buric:2021yak,Hadasz2010, Kraus:2017ezw,Cho:2017oxl,Alkalaev:2017bzx,Rosenhaus:2018zqn,Parikh:2019ygo, Fortin:2019dnq, Goncalves:2019znr, Parikh:2019dvm, Fortin:2019zkm, Alkalaev:2020yvq,Hoback:2020pgj,Fortin:2020zxw, Fortin:2020yjz,Fortin:2020bfq,Haehl2020, Alkalaev:2022kal, Fortin:2023xqq, Alkalaev:2023evp,Pavlov:2023asi, Ammon:2024axd, Ammon:2025cdz} for other works on computing global conformal blocks).
This approach is particularly effective for higher-genus blocks because it is not known how to generalize the conventional Casimir-equation construction to higher-genus surfaces. For higher-genus blocks, the conformal cross-ratios and moduli $\vec x$ are understood as plumbing parameters in a fixed plumbing frame.
This frame is specified by a pair-of-pants decomposition, local canonical coordinates on the three-punctured spheres, and Möbius maps between paired boundary components.
From this result, the conformal block in a different conformal frame can be obtained by the standard coordinate redefinitions which change the parametrization of the moduli and introduces an additional coordinate dependent prefactor.

Possible further extensions, which are not treated in this work, include proving exponentiation of semiclassical conformal blocks in boundary/interface conformal field theories or for semiclassical vacuum blocks (where one or multiple internal operators are given by the identity).
In fact, the latter case cannot be proven using a naive application of the standard oscillator formalism since the oscillator wavefunctions do not exponentiate, even when the conformal block itself seems to exponentiate in the first few orders of an expansion in cross-ratios and moduli.
This is likely due to the oscillator formalism not taking into account the null state $L_{-1}\ket{0}$ and its descendants.
Although one can try to circumvent this by defining a modified oscillator formalism which removes the null states from the start, finding a closed form of both the Virasoro generators and a well-defined inner product on the corresponding quotient Hilbert space with null states removed has proven to be nontrivial.
We therefore leave such a construction for future work.

The outline of this publication is as follows.
We introduce the oscillator formalism and explain how to use it to compute conformal blocks in section~\ref{sec:osc_formalism}.
Global conformal blocks are constructed in the oscillator formalism in section~\ref{sec:global_blocks} while section~\ref{sec:proof_exponentiation} contains the main exponentiation proof.
Some details of the derivations are left for the appendices.
Appendix~\ref{app:diff_eq_deriv} elaborates on the oscillator formalism for three-point vertices of internal lines while appendix~\ref{sec:Kac-determinant} explains how the Kač determinant emerges in the oscillator formalism and proves that the semiclassical conformal block is defined uniquely for non-degenerate internal operators.
Finally, appendix~\ref{sec:semiclassical-vacuum-block} discusses semiclassical vacuum blocks and why our proof does not straightforwardly extend to this case.

\section{The oscillator formalism}
\label{sec:osc_formalism}
In this section, we first review in section~\ref{sec:osc_wavefunctions} the oscillator formalism following \cite{Zamolodchikov1986,Besken:2019bsu} before extending it in section~\ref{sec:osc_wavefcts_three_variables} to oscillator wavefunctions dependent on three oscillator variables and explaining in section~\ref{sec:constr_vir_conf_blocks} how to compute conformal blocks with this method.

In the absence of null states, the Hilbert space of a two-dimensional Euclidean conformal field theory is given by a unitary highest-weight representation of two copies of the Virasoro algebra. Each copy is defined through the commutation relation 
\begin{align} \label{eq:cr_virasoro}
    [L_m,L_n] = (m-n) L_{m+n} + \frac{c}{12}(m^3 - m)\delta_{m+n,0}
\end{align}
for $m,n \in \mathbb{Z}$.
Since we focus on computing conformal blocks, we will need only one of the two copies in the following. A primary field $\cO_h(z)$ defines the corresponding highest-weight state $\ket{h}$ via the operator-state correspondence through $\ket{h} = \cO_h(0) \ket{0}$ and said state fulfills the eigenvalue equation $L_0 \ket{h} = h \ket{h}$ for the dilatation operator $L_0$. The full representation space, the Verma module $\cV_h$, is then freely generated by $L_n$ for $n < 0$ and unitarity restricts us to $c >1 $ and $h > 0$. 

\subsection{Oscillator wavefunctions}
\label{sec:osc_wavefunctions}
A convenient way to introduce an orthogonal inner product on $\cV_h$ is to represent its states in an overcomplete basis of infinitely many monomials $\prod_i u_i^{n_i}$ dependent on so-called \emph{oscillator variables} $u_i$.
A general state $\ket{f}$ in $\cV_h$ is represented by a wavefunction $f(U) = \langle U | f \rangle$, where $U$ denotes the infinite set of oscillator variables.
A suitable inner product between two holomorphic functions $f(U)$ and $g(U)$ can then be shown to be \cite{Zamolodchikov1986}
\begin{align} \label{eq:inner_product}
    (f,g)  = \prod_{n=1}^\infty\int_{\mathbb{C}} \mathrm{d}u_n \mathrm{d}\bar{u}_n \frac{2n}{\pi} \mathrm{e}^{-2nu_n \bar{u}_n} \overline{f(U)}g(U) =: \int_{\mathbb{C}^\infty} \left[\mathrm{d}U\right] \overline{f(U)}g(U)\,.
\end{align}
In this formalism, the projector $\mathbb{P}_h$ that projects general states of the entire CFT Hilbert space into a Verma module of weight $h$ is defined to be
\begin{align}
    \mathbb{P}_h = \int_{\mathbb{C}^\infty} \left[\mathrm{d}U\right] |\overline{U}\rangle\! \langle U |\,.
\end{align}
The inner product \eqref{eq:inner_product} of general basis elements is given by
\begin{align}
    \left(u_1^{n_1} \cdots u_N^{n_N},u_1^{m_1} \cdots u_N^{m_N}\right) = \prod_{k=1}^\infty \frac{n_k!}{(2k)^{n_k}} \delta_{n_k,m_k}\,.
\end{align}

A specific Verma module in this basis is generated by the action of differential operators $l_k^{(U)}$ for $k < 0$ on the highest-weight state $\langle U | h \rangle = 1$ with the differential operator defined through $ \langle U | L_k | \psi \rangle = l_k^{(U)} \langle U | \psi \rangle$ for all $\psi \in \cV_h$.
Using the notation $c = 1+24\mu^2, h = \mu^2+\lambda^2$ for the central charge and conformal weight of the Verma module, these differential operators are given by
\begin{equation} \label{eq:generators_oscillator_basis}
  \begin{aligned}
    l_0^{(U)} &= h + \sum_{n=1}^\infty n u_n \partial_{u_n},\\
    l_{k > 0}^{(U)} &= \sum_{n=1}^\infty n u_n \partial_{u_{n+k}} - \frac{1}{4} \sum_{n=1}^{k-1} \partial_{u_n}\partial_{u_{k-n}} + (\mu k + i \lambda)\partial_{u_k},\\
    l_{-k < 0}^{(U)} &= \sum_{n=1}^\infty (n+k)u_{n+k}\partial_{u_n} - \sum_{n=1}^{k-1}n(k-n)u_nu_{k-n}+2k(\mu k-i\lambda)u_k\,.
  \end{aligned}
\end{equation}
They fulfill the Virasoro algebra from equation \eqref{eq:cr_virasoro},
\begin{equation}
    [l_m^{(U)}, l_n^{(U)}] = (m-n)l_{m+n}^{(U)} + \frac{c}{12}(m^3-m)\delta_{m+n,0}\,.
\end{equation}
A separate set of differential operators $\bar l_k^{(\overline{U})}$ is defined from \eqref{eq:generators_oscillator_basis} by the replacements $u_n \to \bar{u}_n$ and $\lambda \to - \lambda$.\footnote{These operators are not to be confused with the antiholomorphic Virasoro generators.}
These operators act as $\langle \psi |L_k|\overline{U}\rangle = \bar{l}_{-k}^{(\overline{U})} \langle \psi | \overline{U} \rangle$.

To use this basis for the computation of conformal blocks, we first have to determine the different matrix elements of primary operators. More specifically, the oscillator wavefunctions introduced for that purpose in \cite{Zamolodchikov1986,Besken:2019bsu} are defined as
\begin{gather}
    \psi(z_1,z_2,U) = \langle U | \cO_{h_1}(z_1) \cO_{h_2}(z_2) | 0 \rangle \,, \quad \chi(z_1,z_2,\overline{U}) = \langle 0 | \cO_{h_1}(z_1) \cO_{h_2}(z_2) | \overline{U} \rangle\,, \\ \Omega(z_1, U, \overline{V}) = \langle U | \cO_h(z_1) | \overline{V} \rangle\,. \nonumber
\end{gather}
Only the special cases $\psi(z,U) = \langle U|O_h(z)|0 \rangle$ and $\chi(z,\overline{U}) = \langle 0|O_h(z)|\overline{U} \rangle$ are known in closed form for general values of the central charge and conformal dimension \cite{Besken:2019bsu},
\begin{align}
    \psi(z,U) = \exp \left(2(\mu - i \lambda) \sum_{n=1}^\infty z^n u_n\right), \quad \chi(z,U) = z^{-2h} \overline{\psi(z^{-1},U)}\,.
\end{align}
We can however derive a set of differential equations satisfied by $\psi(z_1,z_2,U)$, $\chi(z_1,z_2,\overline{U})$ and $\Omega(z_1, U, \overline{V})$ using equation \eqref{eq:generators_oscillator_basis}. For example, we use for $\psi$ that \cite{Besken:2019bsu}
\begin{align}
    \langle U | L_k \cO_{h_1}(z_1) \cO_{h_2}(z_2) - [L_k, \cO_{h_1}(z_1) \cO_{h_2}(z_2)] | 0 \rangle = 0\,, \quad k \geq -1 
\end{align}
together with the fact that the Virasoro generators act on the primary fields as
\begin{align}
    [L_k, \cO_{h}(z)] = - \cL_k \cO_{h}(z)\,, \qquad \cL_k^{(z)} = -z^{k+1}\partial_z - (k+1)hz^k\,.
\end{align}
This implies the equation \cite{Besken:2019bsu}
\begin{align} \label{eq:DE_psi}
    (l_k^{(U)} + \cL_k^{(z_1)} + \cL_k^{(z_2)}) \psi(z_1,z_2,U) = 0\,, \quad k \geq -1\,.
\end{align}
A very similar computation derives the corresponding \emph{oscillator equations} for $\chi$ as
\begin{align} \label{eq:DE_chi}
    (\overline{l}^{(\overline{U})}_{-k} - \cL_k^{(z_1)} - \cL_k^{(z_2)}) \chi(z_1,z_2,\overline{U}) = 0\,, \quad k \leq 1 \,,
\end{align}
and for $\Omega$
\begin{align} \label{eq:DE_Omega}
    (\cL^{(z)}_k + l_k^{(U)} - \bar{l}_{-k}^{(\overline{V})}) \Omega(z,U,\overline{V}) = 0\,, \quad k \in \ZZ \,.
\end{align}

Although a closed form expression for these wavefunctions is not available, they can be computed order by order in an expansion in monomials,
\begin{equation}
    \psi(z_1,z_2,U) = \psi_{\{\}} + \psi_{\{1\}} u_1 + \left(\psi_{\{1,1\}} u_1^2 + \psi_{\{2\}} u_2\right) + \left(\psi_{\{1,1,1\}} u_1^3 + \psi_{\{2,1\}} u_2 u_1 + \psi_{\{3\}} u_3\right) + \dots\,.
    \label{eq:expansion-monomials}
\end{equation}
By plugging this ansatz into the differential equation \eqref{eq:DE_psi} and grouping together terms of the same level one can solve for the coefficients $\psi_{\{j_1,j_2,\dots\}}$.
In practice, it is often useful to scale out the dependence on the position of the operator insertions in the wavefunction.
For example for the $\psi(z,0,u)$ wavefunction, this is accomplished by scaling out an overall prefactor and introducing a new set of rescaled oscillator variables $\eta_m = z^m u_m$ \cite{Besken:2019jyw},
\begin{equation} \label{eq:rescaling_psi}
    \psi(z,0,u) = z^{h-h_1-h_2}F_\psi(\eta).
\end{equation}
In terms of the new variables, the wavefunction obeys the differential equations\footnote{The $k=0$ and $k=-1$ equations in \eqref{eq:DE_psi} are automatically fulfilled by the ansatz \eqref{eq:rescaling_psi}.}
\begin{equation}
    l_k^{(\eta)}F_\psi(\eta) = \left(l_0^{(\eta)} - h_2 + kh_1\right)F_\psi(\eta), \quad (k \geq 1).
    \label{eq:coordinate-free-DE-psi}
\end{equation}
Expanding $F_\psi(\eta)$ into monomials and gathering terms of the same level $F_\psi = \sum_n F_{\psi,n}$ where $F_{\psi,n} = \sum_i F_{\{i_1,\dots i_p\}} u_{i_p} \dots u_{i_1}$ with $\sum_k i_k = n$ and $i_p \leq i_{p-1} \leq \dots \leq i_1$ gives
\begin{equation}
    l_k^{(\eta)} F_{\psi,n}(\eta) = \left(l_0^{(\eta)} - h_2 + kh_1\right)F_{\psi,n-k}(\eta) = \left(n-k + h - h_2 + k h_1\right)F_{\psi,n-k}(\eta).
\end{equation}
Applying this equation recursively yields \cite{Besken:2019jyw}
\begin{equation}
  M \vec F_{\psi} = \vec B_{\psi},
  \label{eq:coeff-F}
\end{equation}
where $\vec F_\psi$ is a vector of the coefficients $F_{\{i_1,\dots i_p\}}$.
The matrix $M$ is determined by applying the oscillator representations of the Virasoro generators onto a monomial,
\begin{equation}
  M_{ij} = l_{i_p}\dots l_{i_1} u_{j_q}\dots u_{j_1}
  \label{eq:definition-matrix-M}
\end{equation}
while the elements of the vector $B_\psi$ are given by
\begin{equation}
  B_i = \beta_{i_p,i_p}\beta_{i_{p-1},i_p+i_{p-1}}\dots\beta_{i_2,n-i_1}\beta_{i_1,n}
\end{equation}
with $\beta_{k,n} = n-k+h-h_2+kh_1$.
Therefore, the solution of the differential equation reduces to the solution of a linear system of equations.
Similar relations apply to the other wavefunctions, see appendix~\ref{sec:Kac-determinant}.

\subsection{Oscillator wavefunctions with three oscillator variables}
\label{sec:osc_wavefcts_three_variables}
An additional necessary ingredient for the construction of general Virasoro conformal blocks is a generalization of $\Omega(z_1, U, \overline{V}) = \langle U | \cO_h(z_1) | \overline{V} \rangle$, where instead of the matrix element of a primary, we consider the matrix element $\langle U | V(z) | \overline{W} \rangle$ of a general descendant operator $V(z)$. The wavefunction is then dependent on three oscillator variables $U,V,W$. To disambiguate the different conformal weights of the associated oscillator variables, we will denote them by $h_U,h_V,h_W$ in the following. We define in the following the wavefunctions satisfying this property and derive their oscillator equations.

Let us start by elucidating the definition of these wavefunctions before deriving the differential equations they satisfy.
The basic idea for defining three-point wavefunctions is to transform from the oscillator basis $\ket{U}$ to the standard basis $L_{-n_1}\dots L_{-n_k}\ket{h_U}$ and use the known form of descendant operators in the standard basis as contour integrals over the energy momentum tensor $T(z)$ \cite{Ginsparg:1988ui},
\begin{equation}
    \hat{L}_{-n} \cO(z) = \oint_z \frac{\mathrm d w}{2\pi i}\, \frac{1}{(w-z)^{n-1}} T(w) \cO(z),
    \label{eq:descendant-operator-standard-basis}
\end{equation}
to construct descendant operators in the oscillator basis.
Let us first discuss the analogous basis transformation for descendant states rather than descendant operators.
Using the known overlaps between states in the two bases, one can expand
\begin{equation}
    \bra{U} = \sum_{A,B} G^{AB} \bra{U}L_{-A}\ket{h_U}\bra{h_U}(L_{-B})^\dagger, \quad \ket{\overline{U}} = \sum_{A,B} G^{AB} L_{-A}\ket{h_U}\bra{h_U}(L_{-B})^\dagger\ket{\overline{U}}.
\end{equation}
Here $A,B$ run over a complete basis of Virasoro descendants $L_A = L_{a_1}L_{a_2}\dots$ and $G_{AB} = \bra{h_U}(L_{-A})^\dagger L_{-B} \ket{h_U}$ denotes the Gram matrix with inverse $G^{AB}$.
In this way, for instance the one-point wavefunctions can be written as
\begin{align}
    \psi(z_1,z_2,U) &= \sum_{A,B} G^{AB} \bra{U}L_{-A}\ket{h_U}\bra{h_U}(L_{-B})^\dagger \cO_{h_1}(z_1)\cO_{h_2}(z_2)\ket{0},\\
    \chi(z_1,z_2,\overline{U}) &= \sum_{A,B} G^{AB} \bra{h_U}(L_{-B})^\dagger\ket{\overline{U}}\bra{0}\cO_{h_1}(z_1)\cO_{h_2}(z_2)L_{-A}\ket{h_U}.
\end{align}
For the three-point wavefunction, we define similarly a descendant operator $\hat L_{-V}$ in the oscillator basis as
\begin{equation}
  V(z) := \hat L_{-V} O_{h_V}(z) = \sum_{A,B} G^{AB} \bra{V}L_{-A}\ket{h_V} (\hat L_{-B} \cO_{h_V}(z)) \label{eq:def_desc_op}
\end{equation}
where $\hat L_{-B} \cO_{h_V}(z) := \hat L_{-B_1} \hat L_{-B_2}\dots \cO_{h_V}(z)$ denotes an arbitrary descendant operator in the standard basis.
The three-point wavefunction is then defined by
\begin{align} \label{eq:def_xi_canonical}
    \Xi(z,U,V,\overline{W}) = \langle U| V(z) |\overline{W}\rangle = \sum_{A,B} G^{AB} \langle V |L_{-A}| h_V\rangle \langle U |(\hat L_{-B} \cO_{h_V}(z))| \overline{W} \rangle\,.
\end{align}

In order to work directly in the oscillator basis without having to transform to the standard basis, we are interested in deriving differential equations of the form
\begin{equation}
\label{eq:DE_Xi}
  (l_n^{(U)} - \bar{l}_{-n}^{(\overline{W})} + k_n^{(z,V)})\Xi(z,U,V,\overline{W}) = 0\,.
\end{equation}
These arise from inserting operators of the Virasoro algebra and using the commutation relation
\begin{equation}
  [L_n, V(z)] = -k_n^{(z,V)} V(z)
  \label{eq:commutation-relation}
\end{equation}
where $k_n^{(z,V)}$ is an operator in terms of $z,V$ that is to be determined. From the definition \eqref{eq:commutation-relation}, it is clear that the $k_n^{(z,V)}$ obey the centerless Virasoro algebra
\begin{equation}
  [k_m,k_n] = (m-n)k_{m+n}.
  \label{eq:commutator-k}
\end{equation}
where we drop for now the $(z,V)$ superscripts to simplify the notation. Using the linear dilaton theory, we prove in appendix~\ref{app:diff_eq_deriv} that
\begin{equation}
  \begin{aligned}
    k_n &= \cL_n - (n+1)z^n(l_0-h) - \sum_{k=0}^{n-1} \binom{n+1}{k} z^k l_{-(n-k)}, \quad &n \geq -1,\\
    k_n &= \cL_n - (n+1)z^n(l_0-h) + \sum_{k=1}^\infty (-1)^k \binom{-n-1+k}{k+1} z^{n-k} l_{-k}, \quad &n < -1\,.
  \end{aligned}
  \label{eq:ansatz-for-k-arbitrary-n}
\end{equation}
The first few generators, are, for example, given by 
\begin{equation}
  \begin{aligned}
    k_{-2} &= \cL_{-2} + \sum_{k=1}^\infty \frac{(-1)^k}{z^{2+k}} l_{-k} + \frac{1}{z^2}(l_0 - h)\\
    k_{-1} &= \cL_{-1}\\
    k_0 &= \cL_0 - l_0 + h\\
    k_1 &= \cL_1 - 2z(l_0 - h) - l_{-1}\\
    k_2 &= \cL_2 - 3z^2(l_0 - h) - 3z l_{-1} - l_{-2}\,.
  \end{aligned}
  \label{eq:ansatz-for-k}
\end{equation}
Note that these generate all $k_n$ through the commutation relations \eqref{eq:commutator-k}. We give a few more comments in appendix~\ref{app:diff_eq_deriv} as to how this ansatz is obtained. 

For the construction of general conformal blocks we also need the dual wavefunction to $\Xi$, which we define as\footnote{Note that the form of the descendant operators in the definition can be obtained from a consistency condition via $\lim_{z \to 0} \sum_{A,B} G^{AB} (\hat L_{-B} O_{h_V}(z))\ket{0}\bra{h_V}(L_{-A})^\dagger = \sum_{A,B} G^{AB} L_{-B}\ket{h_V}\bra{h_V}(L_{-A})^\dagger = \mathbb{1}$.}
\begin{align} \label{eq:def_lambda}
    \Lambda(z,U,\overline{V},\overline{W}) = \langle U|\overline{V}(z)|\overline{W}\rangle = \sum_{A,B} G^{AB} \langle h_V| (L_{-A})^\dagger | \overline{V}\rangle \langle U |(\hat{L}_{-B} \cO_{h_V}(z))| \overline{W} \rangle\,.
\end{align} 
Its differential equation is given by equation \eqref{eq:DE_Xi}, just with $l_n^{(V)}$ replaced by $\bar{l}_n^{(\overline{V})}$ in \eqref{eq:ansatz-for-k-arbitrary-n},
\begin{equation}
  \begin{aligned}
    \bar{k}_n &= \cL_n - (n+1)z^n(\bar{l}_0-h) - \sum_{k=0}^{n-1} \binom{n+1}{k} z^k \bar{l}_{-(n-k)}, \quad &n \geq -1,\\
    \bar{k}_n &= \cL_n - (n+1)z^n(\bar{l}_0-h) + \sum_{k=1}^\infty (-1)^k \binom{-n-1+k}{k+1} z^{n-k} \bar{l}_{-k}, \quad &n < -1\,,
  \end{aligned}
\end{equation}
from which we find
\begin{equation}
\label{eq:DE_Lambda}
  (l_n^{(U)} - \bar{l}_{-n}^{(\overline{W})} + \bar{k}_n^{(z,\overline{V})})\Lambda(z,U,\overline{V},\overline{W}) = 0.
\end{equation}
The differential equations \eqref{eq:DE_psi}, \eqref{eq:DE_chi}, \eqref{eq:DE_Omega}, \eqref{eq:DE_Xi} and \eqref{eq:DE_Lambda} are the input for the exponentiation proof in section~\ref{sec:proof_exponentiation}.

\subsection{Construction of Virasoro conformal blocks}
\label{sec:constr_vir_conf_blocks}

We now explain how the oscillator wavefunctions introduced in sections \ref{sec:osc_wavefunctions} and \ref{sec:osc_wavefcts_three_variables} are combined into Virasoro conformal blocks by a sewing procedure. The construction is naturally formulated in the \emph{plumbing frame}. Consider a general correlator associated to a Riemann surface $\Sigma_{g,n}$ of genus $g$ with $n$ punctures. This Riemann surface can be decomposed into three-punctured spheres by a pair-of-pants decomposition. In the plumbing frame, each three-punctured sphere is equipped with local canonical coordinates $\xi$ in which the punctures are placed at $\infty$, $1$, $0$, and paired boundary components are related by Möbius maps. The trivalent graph associated with the pair-of-pants decomposition has $e = 3g - 3 + n$ internal edges.\footnote{Except for the trivial cases of the two- and three-point functions on the sphere and the torus partition function. However, for these cases exponentiation holds automatically.} This agrees with the complex dimension of the corresponding moduli space, such that one modulus $x_e$ is associated to each sewn edge $e$. Following standard nomenclature, we refer to these moduli as \emph{plumbing parameters}.\footnote{Regarding terminology, we distinguish here between ``plumbing'' on the level of the conformal structure of a corresponding Riemann surface, and ``sewing'' as the construction of conformal blocks from wavefunctions via the oscillator formalism.}

On the level of wavefunctions, this is realized as follows. We first take each oscillator wavefunction in the local canonical coordinates with punctures at $\infty,1,0$. More concretely, they are given as \begin{gather} 
    \chi(\overline{U};h) := \braket{h}{\overline{U}}\,, \qquad \psi(U;h) := \braket{U}{h}\,, \notag\\
    \chi(\overline{U};h_1,h_2) := \langle h_1 | \cO_{h_2}(1)| \overline{U} \rangle\,, \qquad \psi(U;h_1,h_2) := \langle U |\cO_{h_1}(1)| h_2 \rangle\,, \notag \\ \Omega(U^{(1)},\overline{U}^{(2)};h_1) := \langle U^{(1)} |\cO_{h_1}(1)| \overline{U}^{(2)}\rangle \label{eq:wavefcts_local_coords} \\
    \Xi(U^{(1)},U^{(2)},\overline{U}^{(3)}) := \langle U^{(1)} | U^{(2)}(1) | \overline{U}^{(3)} \rangle\,, \qquad \Lambda(U^{(1)},\overline{U}^{(2)},\overline{U}^{(3)}) := \langle U^{(1)} | \overline{U}^{(2)}(1) | \overline{U}^{(3)} \rangle\,. \notag
\end{gather}
Note that we now explicitly denote the dependence on the external conformal weights which was suppressed in section~\ref{sec:osc_formalism} (the dependence on the internal weights is still suppressed as this can be read off from the oscillator arguments). This collection of oscillator wavefunctions suffices for computing any Virasoro block, as the construction by oscillator wavefunctions is equivalent to the canonical Gram-matrix definition of conformal blocks. After choosing a pair-of-pants decomposition, every three-punctured sphere is represented by one of these ingredients. The following graphical notation helps to keep track of how their legs are sewn:
\begin{gather}
     \chi(\overline{U};h) = \def\tkzscl{0.75}
\begin{tikzpicture}[baseline={([yshift=-1ex]current bounding box.center)},vertex/.style={anchor=base,
     circle,fill=black!25,minimum size=18pt,inner sep=2pt},scale=\tkzscl]
         \coordinate[label=left:$h$\,] (z) at (2.5,0);
         \coordinate[label=right:\,$\overline{U}\phantom{|}$] (u) at (4,0);
         \draw[thick,dashed] (4,0) --node[above] {} (z);
         \filldraw [fill=white] (u) circle (3pt);
        \filldraw (u) circle (1pt);
        \draw (u) circle (3pt);
         \fill (z) circle (3pt);
\end{tikzpicture}\,, \qquad \psi(U;h) = \def\tkzscl{0.75}
\begin{tikzpicture}[baseline={([yshift=-1ex]current bounding box.center)},vertex/.style={anchor=base,
     circle,fill=black!25,minimum size=18pt,inner sep=2pt},scale=\tkzscl]
         \coordinate[label=left:$\phantom{|}U$\,] (u) at (2.5,0);
         \coordinate[label=right:\,$h$] (z) at (4,0);
         \draw[thick,dashed] (4,0) --node[above] {} (u);
         \fill[white] (u) circle (3pt);
         \draw(u) circle (3pt);
         \fill (z) circle (3pt);
     \end{tikzpicture}  \notag \\
    \chi(\overline{U};h_1,h_2) = \def\tkzscl{0.25}
\begin{tikzpicture}[baseline={([yshift=-.5ex]current bounding box.center)},vertex/.style={anchor=base,
    circle,fill=black!25,minimum size=18pt,inner sep=2pt},scale=\tkzscl]
        \coordinate[label=left:$h_2$\,] (z_1) at (-2,2);
        \coordinate[label=left:$h_1$\,] (z_2) at (-2,-2);
        \coordinate[label=right:\,$\overline{U}$] (u_1) at (3,0);
        \draw[thick] (z_1) -- (0,0);
        \draw[thick] (z_2) -- (0,0);
        \draw[thick, dashed] (0,0) -- node[above] {
        } (u_1);
        \fill (z_1) circle (8pt);
        \fill (z_2) circle (8pt);
        \filldraw [fill=white] (u_1) circle (8pt);
        \filldraw (u_1) circle (3pt);
        \draw (u_1) circle (8pt);
\end{tikzpicture}\,, \qquad \psi(U;h_1,h_2) = \def\tkzscl{0.25}
\begin{tikzpicture}[baseline={([yshift=-.5ex]current bounding box.center)},vertex/.style={anchor=base,
    circle,fill=black!25,minimum size=18pt,inner sep=2pt},scale=\tkzscl]
        \coordinate[label=right:\,$h_1$] (z_1) at (2,2);
        \coordinate[label=right:\,$h_2$] (z_2) at (2,-2);
        \coordinate[label=left:$U$\,] (u_1) at (-3,0);
        \draw[thick] (z_1) -- (0,0);
        \draw[thick] (z_2) -- (0,0);
        \draw[thick, dashed] (0,0) -- node[above] {
        } (u_1);
        \fill (z_1) circle (8pt);
        \fill (z_2) circle (8pt);
        \fill[white] (u_1) circle (8pt);
        \draw (u_1) circle (8pt);
\end{tikzpicture}  \notag\\
    \Omega(U^{(1)},\overline{U}^{(2)};h) = \def\tkzscl{0.25}
\begin{tikzpicture}[baseline={([yshift=-.5ex]current bounding box.center)},vertex/.style={anchor=base,
    circle,fill=black!25,minimum size=18pt,inner sep=2pt},scale=\tkzscl]
        \coordinate[label=left:$h_1$\,] (z_1) at (0,3);
        \coordinate[label=left:$U^{(1)}$\,] (u_1) at (-3,0);
        \coordinate[label=right:\,$\overline{U}^{(2)}$] (u_2) at (3,0);
        \draw[thick] (z_1) -- (0,0);
        \draw[thick, dashed] (u_2) --node[above] {
        } (0,0);
        \draw[thick, dashed] (0,0) -- node[above] {
        } (u_1);
        \fill (z_1) circle (8pt);
        \fill[white] (u_1) circle (8pt);
        \draw (u_1) circle (8pt);
        \fill[white] (u_2) circle (8pt);
        \filldraw (u_2) circle (3pt);
        \draw (u_2) circle (8pt);
        \draw[draw=none] (0,0) -- (0,-2.4);
\end{tikzpicture} \label{eq:diagrams_wavefunctions}\\
    \Xi(U^{(1)}, U^{(2)}, \overline{U}^{(3)}) = \def\tkzscl{0.25}
\begin{tikzpicture}[baseline={([yshift=-.5ex]current bounding box.center)},
    vertex/.style={anchor=base,circle,fill=black!25,minimum size=18pt,inner sep=2pt},
    scale=\tkzscl]

    \coordinate (c) at (0,0);

    \def\r{3}

    \coordinate[label=left:$U^{(2)}$\,] (z_1) at ($(c) + (120:\r)$);
    \coordinate[label=left:$U^{(1)}$\,] (z_2) at ($(c) + (-120:\r)$);
    \coordinate[label=right:\,$\overline{U}^{(3)}$] (u_1) at ($(c) + (0:\r)$);

    \draw[thick,dashed] (z_1) -- node[pos=0.35,below] {
    } (c);
    \draw[thick,dashed] (z_2) -- node[pos=0.75,below] {
    } (c);
    \draw[thick,dashed] (c) -- node[pos=0.75,above] {
    } (u_1);

    \filldraw[fill=white] (z_1) circle (8pt);
    \filldraw[fill=white] (z_2) circle (8pt);
    \filldraw[fill=white] (u_1) circle (8pt);
    \filldraw (u_1) circle (3pt);
    \draw (u_1) circle (8pt);

\coordinate (c) at (0,0);

\node at ($(c) + (45:0.75)$) {};

\end{tikzpicture}\,, \qquad 
   \Lambda(U^{(1)},\overline{U}^{(2)},\overline{U}^{(3)}) =  \def\tkzscl{0.25}
\begin{tikzpicture}[baseline={([yshift=-.5ex]current bounding box.center)},
    vertex/.style={anchor=base,circle,fill=black!25,minimum size=18pt,inner sep=2pt},
    scale=\tkzscl]

    \coordinate (c) at (0,0);

    \def\r{3}

    \coordinate[label=left:$U^{(1)}$] (u_1) at ($(c) + (180:\r)$);

    \coordinate[label=right:$\overline{U}^{(2)}$] (z_1) at ($(c) + (60:\r)$);
    \coordinate[label=right:$\overline{U}^{(3)}$] (z_2) at ($(c) + (-60:\r)$);

    \draw[thick,dashed]
      (z_1) -- node[pos=0.35,below] {
      } (c);

    \draw[thick,dashed]
      (z_2) -- node[pos=0.75,below] {
      } (c);

    \draw[thick,dashed]
      (c) -- node[pos=0.75,above] {
      } (u_1);

    \filldraw[fill=white] (u_1) circle (8pt);
    \filldraw[fill=white] (z_2) circle (8pt);
    \filldraw (z_2) circle (3pt);
    \filldraw[fill=white] (z_1) circle (8pt);
    \filldraw (z_1) circle (3pt);

\coordinate (c) at (0,0);

\node at ($(c) + (120:0.75)$) {};

\end{tikzpicture} \notag
\end{gather}
Solid lines denote external primaries, labeled only by their weight $h$, while dashed lines denote both holomorphic and antiholomorphic oscillator variables with corresponding weights.

Before sewing these wavefunctions to a Virasoro block, we first have to ensure the right dependence on the moduli of the full surface. For this, we refer back to equation \eqref{eq:rescaling_psi}, which tells us that the oscillator wavefunctions depend on the coordinates by a rescaling of the oscillator variables (this is up to a weight-dependent prefactor which we absorb into the overall normalization of the block). Consequently, rescaling each oscillator variable associated to an edge $e$ with the plumbing parameter $x_e$ ensures that the conformal block gives the right coordinate dependence and ensures that it is given by a power series in said plumbing parameters. The legs of the oscillator wavefunctions \eqref{eq:wavefcts_local_coords} are sewn together using the inner product \eqref{eq:inner_product}, where for a given edge $e$ the holomorphic oscillator variable $U^{(e)}$ is paired with the antiholomorphic variable $\overline{U}^{(e)}$. Since the inner product pairs only equal levels, the factor $x_e^k$ may be assigned to either side of the sewn edge, or distributed symmetrically between the two sides. The following two examples are therefore equivalent
\begin{equation}
   \int [\mathrm{d}U^{(e)}] \Psi(x_e \overline{U}^{(e)}) \Psi'(U^{(e)}) = \int [\mathrm{d}U^{(e)}] \Psi(\overline{U}^{(e)}) \Psi'(x_e U^{(e)})\,.
\end{equation}
$\Psi$ and $\Psi'$ denote here any oscillator wavefunction and $x_e U$ is a rescaling of every individual $u_k$ variable in $U$, i.e. $u^{(e)}_{k} \mapsto x_e^k u^{(e)}_{k}$. We will in the following group the plumbing parameters with the antiholomorphic variables.

Finally, to construct a general conformal block $\cF$ for a Riemann surface $\Sigma_{g,n}$, we construct the trivalent graph of its pair-of-pants decomposition using the diagrams in \eqref{eq:diagrams_wavefunctions}. The corresponding integral is sewing the legs of the corresponding wavefunctions together using \eqref{eq:inner_product}, after rescaling the oscillator variables of every internal edge
\begin{equation}
    \cF = \int [\mathrm{d}U^{(1)}]\dots\int [\mathrm{d}U^{(3g-3+n)}] \Psi^{(1)} \dots \Psi^{(2g-2+n)}\,,
    \label{eq:conformal-block-oscillator-integral}
\end{equation}
where we kept the rescaling implicit and used that $\Sigma_{g,n}$ can be decomposed into $2g-2+n$ three-punctured spheres.

Lastly, we comment on the conformal frame dependence. As noted in \cite{Cho:2017oxl}, the connection between the plumbing frame and other conformal frames is nontrivial, because descendants do not transform covariantly under the respective conformal map. This can lead to a different parametrization of the moduli and additional coordinate-dependent prefactors, including the conformal anomaly. Both do not restrict the generality of the exponentiation proof in section \ref{sec:proof_exponentiation}. Thus we can absorb all prefactors into the normalization. 

Said prefactors can be reconstructed by using the Möbius transform
\begin{align}
    \xi = \frac{(z-c)(b-a)}{(z-a)(b-c)}
    \label{eq:map_embedding}
\end{align}
whose inverse maps $(\infty,1,0)_\xi$ to $(a,b,c)_z$. Depending on the convention used, additional factors of $x_e^{h_e}$ can be added. Thus, for a chosen set of cross-ratios, the reconstruction consists of fixing target-frame local coordinates and computing the induced leg-factors together with internal factors $x_e^{h_e}$. The next section illustrates this construction for global conformal blocks. 

\section{Construction of global conformal blocks}
\label{sec:global_blocks}
In this section, we discuss how to apply the oscillator formalism to global conformal blocks, including cases where three-point wavefunctions are needed.
We illustrate the construction with two examples.

In the case of the global algebra, the set of oscillator variables reduces to one variable each. The oscillator formalism implements a unitary irreducible representation of the global subalgebra on (square-integrable) holomorphic functions over the complex unit disk $\mathbb{D}$. Following \cite{Besken:2019bsu,Ammon:2024axd}, we choose the inner product on this Hilbert space to be given by
\begin{equation}
    (f,g) = \frac{2h_u-1}{2\pi}\int_\mathbb{D}\frac{\mathrm{d}^2 u}{(1-u\bar{u})^{2-2h_u}}\, \overline{f(u)}g(u) := \int_\mathbb{D} [\mathrm{d} u]\, \overline{f(u)}g(u)\,.
\end{equation}
Note that this differs from the inner product we used in the Virasoro case.
The monomials $\phi_m(u) = u^m$ establish an orthogonal basis for $h_u>1/2$ and their inner product gives 
\begin{equation}
    (u^m,u^n) = \frac{n!}{(2h_u)_n}\,\delta_{m,n}\,.
\end{equation}
Here, $(2h_u)_n = \frac{\Gamma(2h_u+n)}{\Gamma(2h_u)}$ is the Pochhammer symbol.
In this convention, the differential operators $l_n^{(u)}$ implementing the global conformal group in oscillator variables are given by \cite{Besken:2019bsu}
\begin{equation}
    l_n^{(u)} = u^{1-n} \partial_u + (1-n)h_u u^{-n}, \quad \bar{l}_n^{(\bar{u})} = \bar{u}^{1-n}\partial_{\bar{u}} + (1-n)h_u\bar{u}^{-n},
    \label{eq:diff-operators-global}
\end{equation}
where $-1 \leq n \leq 1$.
The oscillator wavefunctions for the global conformal group are given by \cite{Besken:2019bsu}
\begin{equation}
    \begin{aligned}
        \psi(z_1,z_2,u) &= \frac{(z_1-z_2)^{h_u-h_2-h_1}}{(1-z_1u)^{h_u-h_2+h_1}(1-z_2u)^{h_u+h_2-h_1}},\\
        \chi(z_1,z_2,\bar{u}) &= \frac{(z_1-z_2)^{h_u-h_2-h_1}}{(\bar{u}-z_1)^{h_u-h_2+h_1}(\bar{u}-z_2)^{h_u+h_2-h_1}},\\
        \Omega(z_1,u,\bar{v}) &= (z_1-\bar{v})^{h_u-h_1-h_v}(1-z_1u)^{h_v-h_1-h_u}(1-u\bar{v})^{h_1-h_u-h_v}.
    \end{aligned}
\end{equation}
These wavefunctions obey the same differential equations \eqref{eq:DE_psi}, \eqref{eq:DE_chi} and \eqref{eq:DE_Omega} as in the Virasoro case, just with the differential operators \eqref{eq:generators_oscillator_basis} generating the Virasoro group replaced by \eqref{eq:diff-operators-global} due to the different inner product chosen for the global case.

The global wavefunction $\Xi(z,u,v,\bar{w})$ is therefore a solution of 
\begin{equation}
  (l_n^{(u)} - \bar{l}_{-n}^{(\bar{w})} + k_n^{(z,v)})\Xi = 0
\end{equation}
for $n \in \{-1,0,1\}$ where $k_n^{(z,v)} = \cL_n - (n+1)z^n(l_0^{(v)}-h_v) - \frac{1}{2}n(n+1) l_{-1}^{(v)}$ as in \eqref{eq:ansatz-for-k-arbitrary-n}. However, it is in this specific case more effective to directly evaluate the definition in equation \eqref{eq:def_xi_canonical}
\begin{align} \label{eq:def_global_xi}
    \Xi(z,u,v,\bar{w})  = \sum_{a,b \geq 0} G^{ab} \bra{v}(L_{-1})^a\ket{h_v} \bra{u}(\hat{L}_{-1}^b \cO(z))\ket{\bar{w}}\,.
\end{align}
For the global subgroup one has
\begin{gather} 
\begin{split}
\bra{v}\left(L_{-1}\right)^a\ket{h_v} = (2h_v)_a v^a\,, \quad  G^{ab} = \frac{\delta_{ab}}{a! (2h_v)_a }\,, \\ \bra{u}(\hat{L}_{-1}^b \cO(z))\ket{\bar{w}} = \left(\partial_z\right)^b \mel{u} {\cO_{h_v}(z)}{\bar{w}}\,. 
\end{split}
\end{gather}
Using $\hat L_{-1}\cO(z) = \partial_z \cO(z)$ equation \eqref{eq:def_global_xi} reduces to the finite translation 
\begin{align}
    \Xi(z,u,v,\bar{w}) = \sum_{a \geq 0} \frac{(v \partial_z)^a}{a!} \mel{u}{\mathcal O_{h_v}(z)}{\bar{w}} = \mathrm{e}^{v\partial_z} \Omega(z,u,\bar{w})\,,
\end{align}
for $\Omega(z,u,\Bar{w}) = \mel{u}{\mathcal O_{h_v}(z)}{\bar{w}}$. This evaluates to
\begin{align}
\begin{split}
    \Xi(z,u,v,\Bar{w}) 
    &= \Omega(z+v,u,\Bar{w})\\
    &= (z+v-\Bar{w})^{h_u - h_v - h_w}(1-zu-uv)^{h_w - h_u - h_v}(1-u\Bar{w})^{h_v-h_u-h_w} \\
    &=: (z+v-\Bar{w})^{\delta_1}(1-zu-uv)^{\delta_2}(1-u\Bar{w})^{\delta_3}
\end{split} 
\end{align} 
The $\Lambda$-wavefunction is obtained analogously,
\begin{align}
    \Lambda(z,u,\bar{v},\bar{w}) = e^{\bar{v}\partial_z}\Omega(z,u,\bar{w}) = (z + \bar{v}-\bar{w})^{\delta_1}(1-zu-u\bar{v})^{\delta_2}(1-u\bar{w})^{\delta_3}\,.
\end{align}
Note that in this case, the defining differential operator $k_n^{(z,v)}$ for $\Xi$ is given as the conjugation $ k_n^{(z,v)} = e^{-v \cL_{-1}} \cL_n e^{v \cL_{-1}}$ and again similarly for $\Lambda$.
The explicit series expansions are given through
\begin{align}
    \Xi(z,u,v,\bar{w}) = z^{\delta_1}\sum_{k,l,m = 0}^\infty \frac{\tau_{k,l,m}(\delta_1, \delta_2, \delta_3)}{k!\, l!\, m!} \left(z^{-1} \bar{w}\right)^k (zu)^l \left(z^{-1} v\right)^m \\
    \Lambda(z,u,\bar{v},\bar{w}) = z^{\delta_1}\sum_{k,l,m = 0}^\infty \frac{\tau_{k,l,m}(\delta_1, \delta_2, \delta_3)}{k!\, l!\, m!} \left(z^{-1} \bar{w}\right)^k \left(z u\right)^l (z^{-1}\bar{v})^m
\end{align}
for the coefficients
\begin{align}
    \tau_{k,l,m}(\delta_1,\delta_2, \delta_3) &= (-1)^m (-\delta_1+k-l)_m \tau_{k,l}(\delta_1,\delta_2,\delta_3) \\
    \tau_{k,l}(\delta_1,\delta_2,\delta_3) &= \sum_{p=0}^{\text{min}[k,l]} \frac{k!\, l!}{p!(k-p)!(l-p)!} (-\delta_1)_{k-p} (-\delta_2)_{l-p}(-\delta_3)_p\,.
\end{align}
The coefficient $\tau_{k,l}$ defines the corresponding series for $\Omega(z,u,w)$, see \cite{Ammon:2024axd,Ammon:2025cdz,Alkalaev:2015fbw}.

We now want to give some explicit examples of conformal blocks computed using the wavefunctions with three oscillator variables. The two simplest diagrams that necessitate the $\Xi$- and $\Lambda$-wavefunction are the 6-point star-channel block on the sphere and the 2-point OPE-channel block on the torus. Following our discussion in section \ref{sec:constr_vir_conf_blocks}, the latter is constructed in the plumbing frame by self-sewing the corresponding wavefunction 
\begin{align}
\begin{split}
    &\begin{tikzpicture}[baseline={([yshift=-0.5ex]current bounding box.center)},vertex/.style={anchor=base,
    circle,fill=black!25,minimum size=18pt,inner sep=2pt},scale=0.5]
    \coordinate[] (c_1) at (2.5,0);
     \draw[thick,dashed] (5,0) circle (30pt);
     \coordinate[] (c_2) at (3.95,0);
     \coordinate[label=below:$h_1$] (w_2) at (1.5,-1);
     \coordinate[label=above:$h_2$] (w_1) at (1.5,1);
     \draw[thick] (w_1) -- node[above] {
     } (2.5,0);
        \draw[thick] (w_2) -- node[below] {
        } (2.5,0);
        \draw[thick,dashed] (2.5,0) --node[above] {$h_u$
        } (c_2);
        \fill (w_1) circle (4pt);
        \fill (w_2) circle (4pt);
        \coordinate[label=right:$h_v$
        ] (h) at (6,0);
    \end{tikzpicture} = \begin{tikzpicture}[baseline={([yshift=-0.5ex]current bounding box.center)},vertex/.style={anchor=base,
    circle,fill=black!25,minimum size=18pt,inner sep=2pt},scale=0.25]
        \coordinate[label=left:$h_2$\,] (z_1) at (-2,2);
        \coordinate[label=left:$h_1$\,] (z_2) at (-2,-2);
        \coordinate[label=right:\,$\overline{u}$] (u_1) at (3,0);
        \draw[thick] (z_1) -- (0,0);
        \draw[thick] (z_2) -- (0,0);
        \draw[thick, dashed] (0,0) -- node[above] {
        } (u_1);
        \fill (z_1) circle (8pt);
        \fill (z_2) circle (8pt);
        \filldraw [fill=white] (u_1) circle (8pt);
        \filldraw (u_1) circle (3pt);
        \draw (u_1) circle (8pt);
\end{tikzpicture} \,\def\tkzscl{0.25}
\begin{tikzpicture}[
    baseline={([yshift=-.5ex]current bounding box.center)},
    vertex/.style={anchor=base,circle,fill=black!25,
    minimum size=18pt,inner sep=2pt},
    scale=\tkzscl
]

    \coordinate (c) at (0,0);
    \def\r{2.5}

    \coordinate[label=left:$u$\,]     (u)    at ($(c)+(180:\r)$);
    \coordinate[label=below:$v$]      (v)    at ($(c)+(-60:\r)$);
    \coordinate[label=above:$\bar v$] (vbar) at ($(c)+(60:\r)$);

    \draw[thick,dashed] (u) -- (c);
    \draw[thick,dashed] (v) -- (c);
    \draw[thick,dashed] (c) -- (vbar);

    \draw[thick,<->,shorten >=8pt,shorten <=8pt]
        (vbar) to[out=20,in=-20,looseness=3] (v);

    \filldraw[fill=white] (u)    circle (8pt);
    \filldraw[fill=white] (v)    circle (8pt);
    \filldraw[fill=white] (vbar) circle (8pt);
    \filldraw (vbar) circle (3pt);
    \draw (vbar) circle (8pt);

\end{tikzpicture} \\ &=\int [\mathrm{d}^2 u ] [\mathrm{d}^2 v]  \chi(x \bar{u};h_1,h_2) \Xi( v,u,q\bar{v}) \\
    &= (1-q)^{-h_u} {}_2F_1[1-h_u, 2h_v - h_u, 2h_v,q]
    {}_2F_1[h_u, h_u + h_1-h_2,2h_u,-x]\,.
\end{split}
\end{align}
We would like to compare this result with the expression of  \cite{Kraus:2017ezw} derived in a different coordinate frame where the primary operators are inserted on the cylinder.
Following the discussion in section~\ref{sec:constr_vir_conf_blocks}, we relate the local canonical coordinates to the global coordinate $z = \exp \tilde z$. The cross-ratio for this channel is given by $x = y-1$ for $y = z_2/z_1$, which vanishes in the limit when the two insertions approach each other. We need to consider a Möbius transform that relates the local chart on which $\chi(x \Bar{u})$ is defined to global $z$ coordinates with punctures $z_i$, i.e. $(\infty,1,0)_\xi \mapsto (z_2,z_1,0)_z$, using the map \eqref{eq:map_embedding}, which in this case becomes.
\begin{align}
    z = \frac{z_1 z_2 \xi}{z_1 \xi - z_{12}}\,.
\end{align}
Including the factor of $\xi^{2h_2}$ in the definition of $\bra{h_2} = \lim_{\xi \to \infty} \xi^{2h_2} \bra{0}O_{h_2}(\xi)$, the Jacobian factors for the map $\xi \to z$ are given by $\bigl(\xi^{-2} \frac{\partial \xi}{\partial z}\bigr)^{h_2}\bigr|_{\xi \to \infty} = \bigl(\frac{z_2}{z_1}z_{12}\bigr)^{-h_2}$ and $\bigl(\frac{\partial \xi}{\partial z}\bigr)^{h_1}\bigr|_{\xi \to 1} = \bigl(\frac{z_1}{z_2}z_{12}\bigr)^{-h_1}$. Together with the additional factors $(\partial z_1/\partial \tilde z_1)^{h_1}(\partial z_2/\partial \tilde z_2)^{h_2} = z_1^{h_1} z_2^{h_2}$  from passing into cylinder coordinates $z_i = \exp \tilde z_i$, this gives the leg factor
\begin{align}
    \cL_{\text{OPE}} = (1/y-1)^{-h_2}(y-1)^{-h_1}
\end{align}
for $y = z_2/z_1$. Including the contributions $x^{h_u} = (y-1)^{h_u}$ and $q^{h_v}$ for the internal lines reproduces the result in \cite{Kraus:2017ezw},\footnote{The result in \cite{Kraus:2017ezw} is given as a product of the one-point block on the torus and the four-point block on the cylinder with additional operator insertions at $\tilde z=\pm\infty$. Keeping track of leg-factors for the latter factor leads to the formula shown here. Ref.~\cite{Kraus:2017ezw} also includes another formula for the two-point block on the torus which is rescaled by an additional factor of $y^{h_2}$.}
\begin{equation}
    \begin{tikzpicture}[baseline={([yshift=-0.5ex]current bounding box.center)},vertex/.style={anchor=base,
    circle,fill=black!25,minimum size=18pt,inner sep=2pt},scale=0.5]
    \coordinate[] (c_1) at (2.5,0);
     \draw[thick,dashed] (5,0) circle (30pt);
     \coordinate[] (c_2) at (3.95,0);
     \coordinate[label=below:$h_1$] (w_2) at (1.5,-1);
     \coordinate[label=above:$h_2$] (w_1) at (1.5,1);
     \draw[thick] (w_1) -- node[above] {
     } (2.5,0);
        \draw[thick] (w_2) -- node[below] {
        } (2.5,0);
        \draw[thick,dashed] (2.5,0) --node[above] {$h_u$
        } (c_2);
        \fill (w_1) circle (4pt);
        \fill (w_2) circle (4pt);
        \coordinate[label=right:$h_v$
        ] (h) at (6,0);
    \end{tikzpicture} \sim
    \begin{aligned}[t]
        &q^{h_v} (y-1)^{h_u-h_1}(1/y-1)^{-h_2} (1-q)^{-h_u} {}_2F_1[1-h_u, 2h_v - h_u, 2h_v,q]\\
    &\times {}_2F_1[h_u, h_u + h_1-h_2,2h_u,-x]\,.
    \end{aligned}
\end{equation}
As a further cross-check, the same conformal block is equivalently obtained from the $\Lambda$ wavefunction,
\begin{equation}
    \begin{tikzpicture}[baseline={([yshift=-0.5ex]current bounding box.center)},vertex/.style={anchor=base,
    circle,fill=black!25,minimum size=18pt,inner sep=2pt},scale=0.5]
    \coordinate[] (c_1) at (2.5,0);
     \draw[thick,dashed] (5,0) circle (30pt);
     \coordinate[] (c_2) at (3.95,0);
     \coordinate[label=below:$h_1$] (w_2) at (1.5,-1);
     \coordinate[label=above:$h_2$] (w_1) at (1.5,1);
     \draw[thick] (w_1) -- node[above] {
     } (2.5,0);
        \draw[thick] (w_2) -- node[below] {
        } (2.5,0);
        \draw[thick,dashed] (2.5,0) --node[above] {$h_u$
        } (c_2);
        \fill (w_1) circle (4pt);
        \fill (w_2) circle (4pt);
        \coordinate[label=right:$h_v$
        ] (h) at (6,0);
    \end{tikzpicture} = \int [\mathrm{d}^2 u ] [\mathrm{d}^2 v]  \psi(u;h_1,h_2) \Lambda( v,x\bar{u},q\bar{v}).
\end{equation}

The second example is the six-point star-channel block,
\begin{align}
    &\begin{tikzpicture}[baseline={([yshift=-.5ex]current bounding box.center)},vertex/.style={anchor=base,
    circle,fill=black!25,minimum size=18pt,inner sep=2pt}, scale=0.5]
        \coordinate (O) at (0,0);
        \coordinate (A) at (-1.31, -0.75);
        \coordinate (B) at (0, 1.5);
        \coordinate (C) at (1.31, -0.75);
        \coordinate[label=below:$h_1$
        ] (z_1) at (-1.31,-2.25);
        \coordinate[label=left:$h_2$
        ] (z_2) at (-2.62,0);
        \coordinate[label=left:$h_4$
        ] (z_3) at (-1.31,2.25);
        \coordinate[label=right:$h_3$
        ] (z_4) at (1.31,2.25);
        \coordinate[label=right:$h_6$
        ] (z_5) at (2.62,0);
        \coordinate[label=below:$h_5$
        ] (z_6) at (1.31,-2.25);
        \draw[thick, dashed] (O) -- node[above] {$h_u\quad$
        } (A);
        \draw[thick, dashed] (O) -- node[right] {$h_v$
        } (B);
        \draw[thick, dashed] (O) -- node[below] {$h_w$
        } (C);
        \draw[thick] (A) -- node[left] {
        } (z_1);
        \draw[thick] (A) -- node[below] {
        } (z_2);
        \draw[thick] (B) -- node[above] {
        } (z_3);
        \draw[thick] (B) -- node[above] {
        } (z_4);
        \draw[thick] (C) -- node[below] {
        } (z_5);
        \draw[thick] (C) -- node[right] {
        } (z_6);
        \fill (z_1) circle (4pt);
        \fill (z_2) circle (4pt);
        \fill (z_3) circle (4pt);
        \fill (z_4) circle (4pt);
        \fill (z_5) circle (4pt);
        \fill (z_6) circle (4pt);
\end{tikzpicture}
 \qquad = \qquad    \begin{tikzpicture}[
    baseline={([yshift=-.5ex]current bounding box.center)},
    scale=0.5,
    line cap=round
]

\def\R{4pt}
\def\r{2pt}

\def\d{1.55}
\def\a{1.35}
\def\b{1.35}

\coordinate (X) at (0,0);

\coordinate (u)    at (-\d,0);
\coordinate (v)    at (0,\d);
\coordinate (wbar) at (\d,0);

\draw[thick,dashed] (X) -- (u);
\draw[thick,dashed] (X) -- (v);
\draw[thick,dashed] (X) -- (wbar);

\filldraw[fill=white] (u) circle (\R);      
\filldraw[fill=white] (v) circle (\R);      
\filldraw[fill=white] (wbar) circle (\R);   
\fill (wbar) circle (\r);

\node[below] at (u)    {$u$};
\node[right] at (v)    {$v$};
\node[below] at (wbar) {$\bar w$};

\begin{scope}[shift={(-3.15,1.45)}, rotate=-25]
    \coordinate (Lj)   at (0,0);
    \coordinate (ubar) at (\b,0);
    \coordinate (h2)   at (-\a,\a);
    \coordinate (h1)   at (-\a,-\a);

    \draw[thick,dashed] (Lj) -- (ubar);
    \draw[thick] (Lj) -- (h2);
    \draw[thick] (Lj) -- (h1);

    \fill (h2) circle (\R);
    \fill (h1) circle (\R);

    \filldraw[fill=white] (ubar) circle (\R);
    \fill (ubar) circle (\r);

    \node[above] at (ubar) {$\bar u$};
    \node[left]  at (h2)   {$h_2$};
    \node[left]  at (h1)   {$h_1$};
\end{scope}

\coordinate (vbar) at (0,2.00);
\coordinate (Tj)   at (0,3.35);
\coordinate (h3)   at (-1.35,4.70);
\coordinate (h4)   at ( 1.35,4.70);

\draw[thick,dashed] (Tj) -- (vbar);
\draw[thick] (Tj) -- (h3);
\draw[thick] (Tj) -- (h4);

\fill (h3) circle (\R);
\fill (h4) circle (\R);

\filldraw[fill=white] (vbar) circle (\R);
\fill (vbar) circle (\r);

\node[left]        at (vbar) {$\bar v$};
\node[above left]  at (h3)   {$h_4$};
\node[above right] at (h4)   {$h_3$};

\begin{scope}[shift={(3.15,1.45)}, rotate=25]
    \coordinate (Rj) at (0,0);
    \coordinate (w)  at (-\b,0);
    \coordinate (h5) at (\a,\a);
    \coordinate (h6) at (\a,-\a);

    \draw[thick,dashed] (w) -- (Rj);
    \draw[thick] (Rj) -- (h5);
    \draw[thick] (Rj) -- (h6);

    \fill (h5) circle (\R);
    \fill (h6) circle (\R);

    \filldraw[fill=white] (w) circle (\R);

    \node[above] at (w)  {$w$};
    \node[right] at (h5) {$h_6$};
    \node[right] at (h6) {$h_5$};
\end{scope}

\end{tikzpicture} \\ 
    &=\int [\mathrm{d}^2 u ] [\mathrm{d}^2 v] [\mathrm{d}^2 w] \chi(x_1 \bar{u};h_1,h_2) \chi(x_2 \bar{v};h_4,h_3) 
    \Xi(u,v,x_3\bar{w})\psi( w;h_6,h_5) \notag \\ 
    &= \sum_{k,l,m=0}^\infty \frac{
        (h_2 - h_1 + h_u)_l
        (h_3 - h_4 + h_v)_{m}
        (h_6 - h_5 + h_w)_{k}
    }{
        (2h_u)_{l}
        (2h_v)_{m}
        (2h_w)_{k} 
    }
    \tau_{k,l,m}(\delta_1,\delta_2,\delta_3)  x_1^l x_2^m x_3^k\,. \notag
\end{align}
To reconstruct the coordinate dependence, one chooses appropriate plumbing parameters $x_1$, $x_2$, $x_3$ and Möbius transforms relating the canonical three-punctured spheres to global coordinates. A choice for the star-channel cross-ratios that is compatible with the OPE limit of each pair is given in \cite{Fortin:2020zxw} as
\begin{align}
    x_1 = z_{12;54}\,, \qquad
    x_2 = z_{15;43}\,, \qquad
    x_3 = z_{56;14}\,.
\end{align}
for $ z_{ij;kl} = z_{ij}z_{kl} z_{il}^{-1}z_{kj}^{-1}$. The leg factor 
\begin{align}
    \mathcal{L}_{\text{star}} = z_{52;1}^{h_1} z_{15;2}^{h_2}  z_{41;3}^{h_3} z_{13;4}^{h_4} z_{51;6}^{h_6} z_{16;5}^{h_5}\,, \qquad z_{ij;k} := \frac{z_{ij}}{z_{ik}z_{jk}}
\end{align}
in \cite{Fortin:2020zxw} is recovered in the way described before, by choosing Möbius maps relating the local canonical coordinates to the global $z$-coordinates and collecting the associated Jacobian factors. One compatible choice of coordinate maps is obtained by assigning the three triples $(z_1,z_2,z_5)_z$, $(z_4,z_3,z_1)_z$ and $(z_1,z_6,z_5)_z$ to the three canonical triples $(\infty,1,0)_\xi$.  Finally, multiplying with $x_1^{h_u} x_2^{h_v} x_3^{h_w}$ reproduces the result in \cite{Fortin:2020zxw}. The construction of higher-genus blocks in general channels proceeds in the way discussed in section~\ref{sec:constr_vir_conf_blocks}.

\section{Proof of exponentiation}
\label{sec:proof_exponentiation}
In this section, we apply the oscillator formalism to give a proof that Virasoro conformal blocks in the semiclassical limit exponentiate in any channel and for any correlation function.
The broad outline of the proof follows the one put forward for the four-point block on the sphere in \cite{Besken:2019jyw}.
We write the conformal blocks as integrals over oscillator variables, show that the oscillator wavefunctions exponentiate in the semiclassical limit and finally apply a saddle point approximation to obtain the conformal blocks.

Moreover, we provide a small improvement to the proof of \cite{Besken:2019jyw}.
While the latter reference gave a proof of exponentiation for generic values of the internal conformal weight only up to a set of isolated points, we also show that this set is empty.
More precisely, the argument in \cite{Besken:2019jyw} was based on proving that a certain linear system of equations admits a unique solution up to a discrete set of parameter values.
In appendix~\ref{sec:Kac-determinant}, we prove that at these values of the internal conformal weights, degenerate representations of the Virasoro algebra appear.
Since for $c>1$, the only degenerate representation is the vacuum representation $h=0$, this establishes that such isolated points do not exist for semiclassical non-vacuum blocks and exponentiation works for any parameter value.

Consider a generic $n$-point conformal block on a Riemann surface of genus $g$.
In the oscillator formalism, this conformal block is constructed according to section \ref{sec:constr_vir_conf_blocks} given by \eqref{eq:conformal-block-oscillator-integral}
As explained in section~\ref{sec:osc_formalism}, the oscillator wavefunctions can be expanded in monomials in the oscillator variables as in \eqref{eq:expansion-monomials} whose expansion coefficients $\psi_{\{i_1,i_2,\dots\}}$ are determined from \eqref{eq:coeff-F}.
From appendix~\ref{sec:Kac-determinant}, it is clear this system of equations admits a unique solution unless the internal conformal weight is that of a degenerate representation of the Virasoro algebra.
Similar expansions hold for oscillator wavefunctions with two or three oscillator variables.

The first step of the exponentiation proof involves showing that the oscillator wavefunctions exponentiate in the semiclassical limit, 
\begin{equation}
    \begin{aligned}
        &\psi(U;h_U,h_1,h_2) = e^{\mu^2 S_\psi(\rho;g_U,g_1,g_2)},\\
        &\chi(\bar{U};h_U,h_1,h_2) = e^{\mu^2 S_\chi(\bar\rho;g_U,g_1,g_2)},\\
        &\Omega(U,\bar{V};h_U,h_1,h_V) = e^{\mu^2 S_\Omega(\rho,\bar{\sigma};g_U,g_1,g_V)}\\
        &\Xi(U,V,\bar{W};h_U,h_V,h_W) = e^{\mu^2 S_\Xi(\rho,\sigma,\bar\tau;g_U,g_V,g_W)},\\
        &\Lambda(U,\bar{V},\bar{W};h_U,h_V,h_W) = e^{\mu^2 S_\Lambda(\rho,\bar\sigma,\bar\tau;g_U,g_V,g_W)},
    \end{aligned}
    \label{eq:exponential-ansatz-wavefunction}
\end{equation}
where $\rho_n = u_n/\mu$, $\sigma_n = v_n/\mu$, $\tau_n = w_n/\mu$ are rescaled oscillator variables, $c = 1 + 24\mu^2$ and we have assumed that all conformal weights scale with the central charge, $h = \mu^2 g$ and $\lambda = \mu\alpha$.
The exponents $S_\psi(\rho),S_\chi(\bar\rho),$ etc.~are computed by inserting the exponential ansatz \eqref{eq:exponential-ansatz-wavefunction} into the differential equation obeyed by the wavefunction and then keeping only the leading term in the $\mu \to \infty$ limit.
For instance, for the $\psi$ wavefunction we find from \eqref{eq:coordinate-free-DE-psi}
\begin{equation}
    l_{k,\text{cl.}}^{(\rho)}[S_\psi] - l_{0,\text{cl.}}^{(\rho)}[S_\psi] - k g_1 + g_2 = 0, \quad (k \geq 1),
    \label{eq:semiclassical-DE-psi}
\end{equation}
where we have defined
\begin{equation}
    \begin{aligned}
        l_{0,\text{cl.}}^{(\rho)}[S] &= g_U + \sum_{n=1}^\infty n \rho_n \frac{\partial S}{\partial \rho_n} \,, \\
        l_{k>0,\text{cl.}}^{(\rho)}[S] &= \sum_{n=1}^\infty n \rho_n \frac{\partial S}{\partial \rho_{n+k}} - \frac{1}{2}\sum_{n=1}^{k-1}\frac{\partial S}{\partial \rho_n}\frac{\partial S}{\partial \rho_{k-n}} + (k+i\alpha_U)\frac{\partial S}{\partial \rho_k}\,,\\
        l_{-k<0,\text{cl.}}^{(\rho)}[S] &= \sum_{n=1}^\infty (n+k)\rho_{n+k}\frac{\partial S}{\partial \rho_n} - \sum_{n=1}^{k-1}n(k-n)\rho_n\rho_{k-n} + 2k(k-i\alpha_U)\rho_k\,.
    \end{aligned}
    \label{eq:semiclassical-oscillator-operators}
\end{equation}
Note that in the derivation of \eqref{eq:semiclassical-DE-psi} we have first removed the dependence on the external operator insertion points as in \eqref{eq:coordinate-free-DE-psi}.\footnote{That is, we define $u_n \to \eta_n = z^n u_n \to \rho_n = \eta_n/\mu$.}
Analogously, for wavefunctions with multiple oscillator variables we find the following equations in the semiclassical limit from \eqref{eq:DE_Omega} and \eqref{eq:DE_Xi} \footnote{See also \eqref{eq:two-oscillators-fixed-level}, \eqref{eq:three-oscillators} for more details on the derivation.}
\begin{equation}
    \begin{aligned}
        -kg_1 - l_{0,\text{cl.}}^{(\rho)}[S_\Omega] + \bar{l}_{0,\text{cl.}}^{(\overline\sigma)}[S_\Omega] + l_{k,\text{cl.}}^{(\rho)}[S_\Omega] - \bar{l}_{-k,\text{cl.}}^{(\overline\sigma)}[S_\Omega] = 0,\\
        - l_{0,\text{cl.}}^{(\rho)}[S_\Xi] - n l_{0,\text{cl.}}^{(\sigma)}[S_\Xi] + \bar{l}_{0,\text{cl.}}^{(\overline\tau)}[S_\Xi] + l_{n,\text{cl.}}^{(\rho)}[S_\Xi] - \bar{l}_{-n,\text{cl.}}^{(\overline\tau)}[S_\Xi] + k_{n,\text{cl.}}^{(\sigma)}[S_\Xi] = 0,
    \end{aligned}
\end{equation}
with
\begin{equation}
    \begin{aligned}
        k_{n,\text{cl.}}^{(\sigma)}[S] &= - \sum_{k=0}^{n-1}\binom{n+1}{k}l_{-(n-k),\text{cl.}}^{(\sigma)}[S], \quad (n \geq -1),\\
        k_{n,\text{cl.}}^{(\sigma)}[S] &= \sum_{k=1}^\infty (-1)^k \binom{-n-1+k}{k+1}l_{-k,\text{cl.}}^{(\sigma)}[S], \quad (n < -1).
    \end{aligned}
\end{equation}
Expanding the exponents $S$ into monomials in the semiclassical oscillator variables $\rho,\sigma,\tau$,
\begin{equation}
    S_\psi(\rho) = S_{\{1\}} \rho_1 + \left(S_{\{1,1\}} \rho_1^2 + S_{\{2\}} \rho_2\right) + \left(S_{\{1,1,1\}} \rho_1^3 + S_{\{2,1\}} \rho_2 \rho_1 + S_{\{3\}} \rho_3\right) + \dots\,.
\end{equation}
allows for solving the differential equations order by order.
As explained in appendix~\ref{sec:Kac-determinant}, there is a unique solution for the coefficients $S_{\{i_1,i_2,\dots\}}$.
This proves that the oscillator wavefunctions exponentiate in the semiclassical limit.

The final step for computing semiclassical conformal blocks involves evaluating the integral over oscillator variables,
\begin{equation}
    \cF = \int [\mathrm dU^{(1)}]\dots\int [\mathrm dU^{(3g-3+n)}] e^{\mu^2(S^{(1)} + \dots + S^{(2g-2+n)})}.
\end{equation}
Changing variables from $u^{(e)}_n$ to $\rho^{(e)}_n = u^{(e)}_n/\mu$ yields (taking into account the integration measure)
\begin{equation}
    \cF = \int \left(\prod_m \mathrm d^2\rho^{(1)}_m \frac{2m\mu^2}{\pi}\right) \dots \int \left(\prod_m \mathrm d^2\rho^{(3g-3+n)}_m \frac{2m\mu^2}{\pi}\right) e^{\mu^2(-2\sum_{m,e} m\rho_m^{(e)}\bar\rho_m^{(e)} + \sum_j S^{(j)})}.
    \label{eq:integral-saddle-point-approximation}
\end{equation}
Because $\mu$ is large, the integrals can be evaluated by saddle point approximation,
\begin{equation}
    \cF = e^{\mu^2 \cI + O(\mu^0)}.
\end{equation}
The saddle point values of $\rho_m^{(e)}$ are determined by
\begin{equation}
    2m\rho_m^{(e)} = \frac{\partial}{\partial\bar\rho_m^{(e)}} \sum_j S^{(j)}, \quad 2m\bar\rho_m^{(e)} = \frac{\partial}{\partial\rho_m^{(e)}} \sum_j S^{(j)}.
\end{equation}
This approximation is valid because there is a unique non-degenerate saddle point.
To see this, consider first the example of the four-point block on the sphere.
In this case, the exponent in \eqref{eq:integral-saddle-point-approximation} is given by
\begin{equation}
    S_\text{tot} = -2\sum_m m \rho_m \bar\rho_m + S_\psi(x\rho) + S_\chi(\bar\rho)
\end{equation}
where $(x\rho)_m = \rho_m x^m$ is the rescaled oscillator variable containing the dependence on the cross-ratio $x$.
The first saddle point equation $2m\rho_m = \frac{\partial S_\chi}{\partial\bar\rho_m}$ can be solved immediately.
Plugging the result back into the exponent, inserting the series expansion $\bar\rho_m = \sum_{k \geq m} \rho_{m,k} x^k$ and expanding the exponent in $x$ gives a set of saddle-point equations for each expansion order.
The $k$-th order in the expansion gives $k$ independent equations for the coefficients $\rho_{m,k}$ with $m \leq k$ and can thus be solved uniquely.
Moreover, the saddle point is non-degenerate: $\det(X_{ij}) \neq 0$ for $X_{ij} = \partial_i \partial_j S_\text{tot}$, $i,j = \rho_m,\bar\rho_m$ as can be seen already from the leading order in the cross-ratio expansion arising from the integration measure term $-2\sum_m \rho_m \bar\rho_m$.
Therefore, use of the saddle point approximation is justified order by order in the cross-ratio expansion.
The same argument works for other conformal blocks.
By expanding in cross-ratios and moduli, the saddle-point equation can be solved uniquely order by order.
The saddle-point is non-degenerate already at leading order due to the term coming from the integration measure.

The above arguments show that the limit \eqref{eq:precise-statement-exponentiation} exists to any order in the series expansion into powers of the conformal cross-ratios and moduli.
Therefore, the semiclassical conformal block is well-defined as a formal power series.
This establishes exponentiation of the conformal block within the radius of convergence of this series expansion, although the existence of a non-zero convergence radius and the analytic structure of the semiclassical conformal block is not determined by the oscillator formalism.

\acknowledgments

The authors would like to thank Martin Ammon for comments on the draft. JH would like to thank Jonah Baerman for discussions and Martin Ammon, Tobias Hössel and Katharina Wölfl for collaboration on related work. Work at VUB is supported by FWO-Vlaanderen projects G012222N and G0A2226N, and by the VUB Research Council through the Strategic Research Program High-Energy Physics. MG is supported by FWO-Vlaanderen through a Junior Postdoctoral Fellowship 1238224N. JH is funded by a \emph{Landesgraduiertenstipendium} of the federal state of Thuringia.

\appendix
\section{Differential equation for the wavefunction with three oscillator variables}
\label{app:diff_eq_deriv}

In this appendix, we explain how the differential equation \eqref{eq:DE_Xi} for the oscillator wavefunction with three oscillator variables is obtained. We first give some intuition as to how we found the ansatz \eqref{eq:ansatz-for-k} for the differential operator $k_n^{(z,U)}$ appearing in \eqref{eq:DE_Xi} before giving a complete derivation.

\subsection{Motivation for the ansatz} 
Consider $k_{-1}$ first.
We want to compute the commutator
\begin{equation}
  [L_{-1}, \hat L_{-n} \mathcal O_h(z)]
  \label{eq:commutator-L-1}
\end{equation}
where $\hat L_{-n} \mathcal O_h$ is given by the contour integral \eqref{eq:descendant-operator-standard-basis} with the integration contour encircling $w=z$ and the integral thus computing the residue at this point.
However, one can equivalently think of the integration contour as encircling the entire complex plane except the point $w=z$.
Since the only other residues in the integrand come from the points $w=0,\infty$, we find
\begin{equation}
  \hat L_{-n} \mathcal O_h(z) = \sum_{m=0}^\infty \binom{n-2+m}{n-2} (z^{-n-m+1}(-1)^n \mathcal O_h L_{m-1} + z^m L_{-n-m} \mathcal O_h).
\end{equation}
Multiplying out the commutators in \eqref{eq:commutator-L-1} and using $[L_n, \mathcal O_h(z)] = - \cL_n \mathcal O_h(z)$, we get
\begin{equation*}
  [L_{-1}, \hat L_{-n} \mathcal O_h(z)] = \sum_{m=0}^\infty  \binom{ n-2+m}{n-2}
  \begin{aligned}[t]
    \bigl[&z^{-n-m+1}(-1)^n (- \cL_{-1} \mathcal O_h L_{m-1} -m \mathcal O_h L_{m-2})\\
    + &z^m ((n+m-1)L_{-n-m-1}\mathcal O_h - L_{-n-m}\cL_{-1}\mathcal O_h)\bigr]\,.
  \end{aligned}
\end{equation*}
Pulling out the $\cL_{-1}$ in front and renaming $m \to m-1$ yields
\begin{equation}
  [L_{-1}, \hat L_{-n} \mathcal O_h(z)] = -\cL_{-1}(\hat L_{-n} \mathcal O_h(z)).
\end{equation}
Repeating this argument for $\hat L_{-n_1}\hat L_{-n_2} \mathcal O_h(z)$, $\hat L_{-n_1}\hat L_{-n_2} \hat L_{-n_3} \mathcal O_h(z)$ and so on justifies the identification
\begin{equation}
  [L_{-1}, \hat L_{-V} \mathcal O_h(z)] = -\cL_{-1} (\hat L_{-V} \mathcal O_h(z)) \quad \Leftrightarrow \quad k_{-1} = \cL_{-1}\,.
\end{equation}

Using the same computation steps, one can show that
\begin{equation}
  \begin{aligned}
    [L_0, \hat L_{-n} \mathcal O_h(z)] &= -(\cL_{0}-n)(\hat L_{-n} \mathcal O_h(z)),\\
    [L_0, \hat L_{-n_1} \hat L_{-n_2} \mathcal O_h(z)] &= -(\cL_{0}-n_1-n_2)(\hat L_{-n_1} \hat L_{-n_2} \mathcal O_h(z))\,.
  \end{aligned}
\end{equation}
Since $l_0-h$ is the operator that determines the level of a given monomial,
\begin{equation}
  (l_0 - h) \prod_k u_k^{n_k} = (\sum_k n_k)\prod_k u_k^{n_k},
\end{equation}
it should be the case that
\begin{equation}
  k_0 = \cL_0 - l_0 + h\,.
\end{equation}

Commuting with other operators gives terms that change the level.
For instance, commuting $\hat L_{-n} \mathcal O_h(z)$ with $L_1$ gives a term $\hat L_{-(n-1)} \mathcal O_h(z)$,
\begin{equation}
  [L_1,\hat L_{-n} \mathcal O_h(z)] = -(\cL_1 -2zn)\hat L_{-n} \mathcal O_h(z) + (n+1) \hat L_{-(n-1)} \mathcal O_h(z).
\end{equation}
Changing the level of a given monomial can be accomplished with $l_{k \neq 0}$, so terms of this form should be identified with such operators.
In this way, one can guess the ansatz \eqref{eq:ansatz-for-k}.

\subsection{Derivation}

In order to prove the ansatz, we will use the linear dilaton theory as a concrete example of a CFT.
Since the operators $k_n$ that we want to derive are fixed by conformal symmetry, the proof then applies to any CFT.
The linear dilaton theory contains a scalar field
\begin{equation}
  \partial X(z)
  =
  -i\sum_{m\in\mathbb Z}\alpha_m z^{-m-1},
  \qquad
  [\alpha_m,\alpha_n]=m\delta_{m+n,0}, \label{eq:dX-mode-expansion}
\end{equation}
in terms of which the energy-momentum tensor is given by
\begin{equation} \label{eq:def_T}
  T(z)
  =
  -\frac12:\partial X(z)\partial X(z):
  -\sqrt2\mu\,\partial^2X(z)\,.
\end{equation}
The Hilbert space is spanned by the usual Fock basis obtained by applying $\alpha_{-n}$ operators onto primary states $\ket{h}$.
These primary states correspond to vertex operators $ V_\kappa(z) = :e^{-\sqrt{2}\kappa X(z)}:$ with $h = -\kappa^2 + 2\mu \kappa$ for $ \kappa = \mu + i \lambda$.\footnote{The vertex operators $V_\kappa(z)$ should not be confused with descendant operator $V(z)$ in the oscillator basis in equation \eqref{eq:def_desc_op}.}
An arbitrary descendant state in the oscillator basis can be expanded in terms of the Fock space basis as
\begin{equation}
  \ket{U} = \sum_{k_1=0}^\infty \frac{(i\sqrt{2}\alpha_{-1}u_1)^{k_1}}{k_1!} \sum_{k_2=0}^\infty \frac{(i\sqrt{2}\alpha_{-2}u_2)^{k_2}}{k_2!} \cdots \ket{h}
  \label{eq:oscillator-state}
\end{equation}
where $\ket{h} = V_{\kappa}(0)\ket{0}$.
Writing out the first few terms of $\ket{U}$, one can easily convince oneself that the consistency condition
\begin{equation}
  L_n \ket{U} = l_{-n}^{(U)}\ket{U}
\end{equation}
is fulfilled.
Here, $l_n$ is defined in \eqref{eq:generators_oscillator_basis} and the Virasoro generators $L_n$ for the linear dilaton theory can be written in terms of modes of $\partial X(z)$,
\begin{equation}
  L_n = \frac{1}{2}\sum_m :\alpha_{n-m}\alpha_m: - i \sqrt{2}\mu(n+1)\alpha_n.
\end{equation}
The operator-state correspondence then suggests that an arbitrary descendant operator in the oscillator basis can be written as
\begin{align}
\begin{split}
  &\hat L_{-U} V_{\kappa}(z) := U(z)\\ &=~ : \sum_{k_1=0}^\infty \frac{1}{k_1!}\left(\frac{-\sqrt{2}\partial X(z) u_1}{0!}\right)^{k_1} \dots \sum_{k_m=0}^\infty \frac{1}{k_m!}\left(\frac{-\sqrt{2}\partial^m X(z) u_m}{(m-1)!}\right)^{k_m} \cdots V_{\kappa}(z):
\end{split}
\end{align}
since $U(0)\ket{0} = \ket{U}$ as defined in \eqref{eq:oscillator-state}.
Now, the task at hand is to compute the commutators
\begin{equation} \label{eq:def_kn_contour_integral}
  - k_n^{(z,U)} U(z) := [L_n, U(z)] = \oint_0 \frac{\mathrm{d}w}{2\pi i} w^{n+1} T(w) U(z).
\end{equation}
Note that the sign on the left-hand side is fixed by the convention
\begin{equation}
  [L_n,\cO_h(z)]
  =
  z^{n+1}\partial_z \cO_h(z)
  +(n+1)hz^n\cO_h(z)
  :=
  -\mathcal L_n \cO_h(z).
\end{equation}
For the linear dilaton theory, the computation of the commutators can be accomplished by simple Wick contractions.
Let us define the following shorthand notation,
\begin{equation}
  \cO_{\{k\}}(z) =~ :\prod_{m=1}^\infty \frac{1}{k_m!} \left(\frac{-\sqrt{2}\partial^m X(z) u_m}{(m-1)!}\right)^{k_m}:
\end{equation}
and
\begin{equation}
  U(z) = \sum_{\{k\}} : \cO_{\{k\}}(z) V_{\kappa}(z) :\,.
\end{equation}
Note that we will in the following evaluate the contour-integral in \eqref{eq:def_kn_contour_integral} for $n \in \mathbb{Z}$, where the results for $n < -1 $ are understood by the usual analytic continuation of the binomial coefficient
\begin{equation}
  \oint_z \frac{\mathrm{d}w}{2\pi i} w^{n+1} \frac{1}{(w-z)^{m+2}} = \left\{
    \begin{aligned}
      & \binom{n+1}{m+1} z^{n-m} \quad & ,n \geq m \geq -1\\
      & 0 \quad & ,-1 \leq n < m\\
      & (-1)^{m+1} \binom{-n+m-1}{m+1} z^{n-m} \quad & ,n < -1\,.
    \end{aligned}
  \right. 
\end{equation}
The elementary Wick contractions that we are going to use are
\begin{gather}
  \contraction{}{\partial X(w)}{}{\partial X(z)}\partial X(w)\partial X(z)
  =
  -\frac{1}{(w-z)^2}\,, \qquad
  \contraction{}{\partial X(w)}{}{\partial^m X(z)}\partial X(w)\partial^m X(z)
  =
  -\frac{m!}{(w-z)^{m+1}}\,, \nonumber\\
  \label{eq:elementary-contractions}
  \contraction{}{\partial^2 X(w)}{}{\partial^m X(z)}\partial^2 X(w)\partial^m X(z)
  =
  \frac{(m+1)!}{(w-z)^{m+2}}\,, \qquad
  \contraction{}{\partial X(w)}{}{V_\kappa(z)}\partial X(w)V_\kappa(z)
  =
  \frac{\sqrt2\kappa}{w-z}V_\kappa(z), 
  \\
  \contraction{}{\partial^2 X(w)}{}{V_\kappa(z)}\partial^2 X(w)V_\kappa(z)
  =
  -\frac{\sqrt2\kappa}{(w-z)^2}V_\kappa(z)\,. \nonumber
\end{gather}

We now evaluate the contour integral in equation \eqref{eq:def_kn_contour_integral} via the residue theorem. Start with the contributions from the background-charge term $-\sqrt{2}\mu \partial^2 X(z)$ in equation \eqref{eq:def_T}, of which there are two.
\begin{enumerate}
    \item First consider the contraction of $\partial^2X(w)$ with
$\mathcal O_{\{k\}}(z)$. Using the above contractions \eqref{eq:elementary-contractions}, one obtains
\begin{equation}
  \contraction{}{\partial^2X(w)}{}{\mathcal O_{\{k\}}(z)}\partial^2X(w)\mathcal O_{\{k\}}(z)
  =
  -\sqrt2
  \sum_{m=1}^{\infty}
  \frac{m(m+1)u_m}{(w-z)^{m+2}}
  \mathcal O_{\{\widetilde k(m)\}}(z),
\end{equation}
where we mean by $\widetilde k(m)$ that
\begin{equation}
  \widetilde k_m=k_m-1,
  \qquad
  \widetilde k_p=k_p
  \quad (p\neq m).
\end{equation}
Therefore
\begin{align}
\begin{split}
  &-\sqrt2\mu
  \oint_z\frac{\mathrm{d}w}{2\pi i}\,
  w^{n+1}
  \contraction{}{\partial^2X(w)}{}{\mathcal O_{\{k\}}(z)}\partial^2X(w) :\mathcal O_{\{k\}}(z) V_\kappa(z):
   \\ &=
  2\mu
  \sum_{m=1}^{\infty}
  m(m+1)
  \binom{n+1}{m+1}
  z^{n-m}
  u_m:
  \mathcal O_{\{\widetilde k(m)\}}(z)V_\kappa(z):\,.
\end{split}
  \label{eq:A1-background-O}
\end{align}
After summing over $\{k\}$ with $k_m \geq 1$, this becomes
\begin{equation}
  2\mu
  \sum_{m=1}^{\infty}
  m(m+1)
  \binom{n+1}{m+1}
  z^{n-m}
  u_m U(z).
  \label{eq:A1-background-O-summed}
\end{equation}
  \item The second contribution is given by the contraction of $\partial^2X(w)$ with the vertex operator
\begin{align}
\begin{split}
  &-\sqrt2\mu
  \oint_z\frac{\mathrm{d}w}{2\pi i}\,
  w^{n+1}
  \contraction{}{\partial^2X(w)}{\mathcal O_{\{k\}}(z)}{V_\kappa(z)}\partial^2X(w):\mathcal O_{\{k\}}(z) V_\kappa(z):
  \\ &=
  2\mu\kappa(n+1)z^n :\mathcal O_{\{k\}}(z) V_\kappa(z):. \end{split}
  \label{eq:A2-background-V}
\end{align}
\end{enumerate}

Next we consider the contractions from the free boson piece $-\frac12:\partial X(z)\partial X(z):$ in equation \eqref{eq:def_T}. Define the contraction of one $\partial X$ with $\cO_{\{k\}}$ by 
\begin{equation}
   \contraction{}{\partial X(w)}{}{\cO_{\{k\}}(z)} \partial X(w) \cO_{\{k\}}(z)
  =
  \sum_{m=1}^{\infty}
  \frac{\sqrt2\,m u_m}{(w-z)^{m+1}}
  \mathcal O_{\{\widetilde k(m)\}}(z)\,.
\end{equation}
These single-contraction terms constitute then the following contributions:
\begin{enumerate}
\item The double contraction of both $\partial X(w)$-factors with
$\mathcal O_{\{k\}}(z)$ gives
\begin{align}
   \contraction{:\partial X(w)}{\partial X(w):}{}{\cO_{k}(z)}
    \contraction[1.5ex]{}{:\partial X(w)}{\partial X(w):}{\cO_{\{k\}}(z)}
    :\partial X(w)\partial X(w):\cO_{\{k\}}(z) &=  2 \sum_{m_1,m_2=1}^{\infty} \frac{m_1 u_{m_1} m_2 u_{m_2}}{(w-z)^{m_1+m_2+2}} \cO_{\{\Tilde{k}(m_1,m_2)\}}(z)\,. 
\end{align}
Taking the contour integral results in
\begin{align}
  &-\frac12
  \oint_z\frac{\mathrm d w}{2\pi i}\,
  w^{n+1}
  \,2
  \sum_{m_1,m_2=1}^{\infty}
  \frac{
    m_1u_{m_1}m_2u_{m_2}
  }{
    (w-z)^{m_1+m_2+2}
  }
  :
  \mathcal O_{\{\widetilde k(m_1,m_2)\}}(z)
  V_\kappa(z):
  \label{eq:B-double-O} \\
  &\qquad =
  -
  \sum_{m_1,m_2=1}^{\infty}
  m_1m_2
  \binom{n+1}{m_1+m_2+1}
  z^{n-m_1-m_2}
  u_{m_1}u_{m_2}
  :
  \mathcal O_{\{\widetilde k(m_1,m_2)\}}(z)
  V_\kappa(z):\,.
  \nonumber
\end{align}
Here $\widetilde k(m_1,m_2)$
means that $k_{m_1}$ and $k_{m_2}$ are reduced by one. If $m_1=m_2$, then $k_{m_1}$ is reduced by two.  After summing over $\{k\}$ with $k_{m_1}, k_{m_2} \geq 1$ ($k_m \geq 2$ for $m_1 = m_2$), we get
\begin{equation}
  -\sum_{m_1,m_2=1}^{\infty}
  m_1m_2
  \binom{n+1}{m_1+m_2+1}
  z^{n-m_1-m_2}
  u_{m_1}u_{m_2}
  U(z).
  \label{eq:B-double-O-summed}
\end{equation}
\item The single contractions with $\cO_{\{k\}}$ give together
\begin{align} 
\begin{split}
  &-\frac12\cdot 2 \sum_{m=1}^{\infty}
  \oint_z\frac{\mathrm{d}w}{2\pi i}\,
  w^{n+1} \frac{\sqrt2\,m u_m}{(w-z)^{m+1}}
  :
  \partial X(w) 
  \mathcal O_{\{\widetilde k(m)\}}(z)  V_\kappa(z)
  :
  \\
  &=
  -\sqrt2
  \sum_{m=1}^{\infty}
  \sum_{r=0}^{m}
  m u_m
  \binom{n+1}{m-r}
  z^{n+1-m+r}
  \frac{1}{r!}
  :
  \partial^{r+1}X(z)
  \mathcal O_{\{\widetilde k(m)\}}(z) V_\kappa(z)
  :\,. \end{split} \label{eq:C-single-O} 
\end{align}
The $r=m$ term is given by
\begin{align}
\begin{split}
  &-\sqrt2
  \sum_{m=1}^{\infty}
  m u_m z^{n+1}
  \frac{1}{m!}
  :
  \partial^{m+1}X(z)
  \mathcal O_{\{\widetilde k(m)\}}(z) V_\kappa(z)
  :\\
  &=
  z^{n+1}:\partial_z\mathcal O_{\{k\}}(z) V_\kappa(z):
  \end{split}
  \label{eq:C2-z-derivative-O}
\end{align}
and consequently contributes to $\cL_n$, as we will see below. For $0\leq r\leq m-1$, we rewrite the insertion of
$\partial^{r+1}X$ as a derivative with respect to $u_{r+1}$
\begin{equation}
  \partial_{u_{r+1}}U(z)
  =
  -\frac{\sqrt2}{r!}
  \sum_{\{k\}}
  :
  \partial^{r+1}X(z)
  \mathcal O_{\{k\}}(z)
 V_\kappa(z)
  : .
  \label{eq:u-derivative-identity}
\end{equation}
Consequently, the remainder of the single-contraction term is
\begin{equation}
  \sum_{m=1}^{\infty}
  \sum_{r=0}^{m-1}
  m u_m
  \binom{n+1}{m-r}
  z^{n+1-m+r}
  \partial_{u_{r+1}}U(z).
  \label{eq:C1-remainder}
\end{equation}
\item The mixed contraction of one $\partial X(w)$ with
$\mathcal O_{\{k\}}(z)$ and the other with $V_\kappa(z)$ is
\begin{align}
  &-\frac12\cdot 2
  \sum_{m=1}^{\infty}
  \oint_z\frac{\mathrm d w}{2\pi i}\,
  w^{n+1}
  \frac{\sqrt2\,m u_m}{(w-z)^{m+1}}
  \frac{\sqrt2\kappa}{w-z}
  :
  \mathcal O_{\{\widetilde k(m)\}}(z) V_\kappa(z):
  \nonumber\\
  &=
  -2\kappa
  \sum_{m=1}^{\infty}
  m u_m
  \binom{n+1}{m+1}
  z^{n-m}
  :
  \mathcal O_{\{\widetilde k(m)\}}(z) V_\kappa(z):
  .
  \label{eq:D-mixed}
\end{align}
\item 
There are now two further contributions coming involving only contractions with $V_\kappa$ and the free boson part. The double contraction of both $\partial X(w)$-factors with $V_\kappa$ contributes as
\begin{align} \label{eq:E1}
    -\kappa^2 (n+1)z^n : \cO_{\{k\}}(z)V_\kappa(z):\,,
\end{align}
while the single contractions, where one $\partial X(w)$ contracts with $V_\kappa$ and the other remains uncontracted, give together 
\begin{align} \label{eq:E2}
    z^{n+1} : \cO_{\{k\}}(z) \partial_z V_\kappa(z):\,.
\end{align}
\end{enumerate}

Let us now combine all possible terms. Note that equations \eqref{eq:A2-background-V}, \eqref{eq:C2-z-derivative-O}, \eqref{eq:E1} and \eqref{eq:E2} amount together exactly to $-\cL_n U(z)$ using $h = -\kappa^2 +2\mu \kappa$. We additionally group \eqref{eq:A1-background-O-summed} and \eqref{eq:D-mixed} together and leave \eqref{eq:B-double-O} and \eqref{eq:C1-remainder} as they are. We thus find in this order
\begin{align}
\begin{split}
      [L_n,U(z)]
  =
  {}&
  -\mathcal L_n U(z)
  \\
  &+
  \sum_{m=1}^{\infty}
  \binom{n+1}{m+1}
  z^{n-m}
  2m\bigl(\mu(m+1)-\kappa\bigr)u_mU(z)
  \\
  &-
  \sum_{m_1,m_2=1}^{\infty}
  \binom{n+1}{m_1+m_2+1}
  z^{n-m_1-m_2}
  m_1m_2u_{m_1}u_{m_2}U(z)
  \\
  &+
  \sum_{m=1}^{\infty}
  \sum_{r=0}^{m-1}
  \binom{n+1}{m-r}
  z^{n+1-m+r}
  m u_m\partial_{u_{r+1}}U(z).
\end{split}
  \label{eq:LnU-general-binomial}
\end{align}
We used here that $\mu(m+1)-\kappa = \mu m-i\lambda$. For $n\geq -1$, the binomial coefficients truncate the sums and this becomes
\begin{align}
\begin{split}
  [L_n,U(z)]
  =
  {}&
  -\mathcal L_n U(z)
  \\
  &+
  \sum_{m=1}^{n}
  \binom{n+1}{m+1}
  z^{n-m}
  2m(\mu m-i\lambda)u_mU(z)
  \\
  &-
  \sum_{m_1=1}^{n-1}
  \sum_{m_2=1}^{n-m_1}
  \binom{n+1}{m_1+m_2+1}
  z^{n-m_1-m_2}
  m_1m_2u_{m_1}u_{m_2}U(z)
  \\
  &+
  \sum_{m=1}^{\infty}
  \sum_{r=\max(0,m-n-1)}^{m-1}
  \binom{n+1}{m-r}
  z^{n+1-m+r}
  m u_m\partial_{u_{r+1}}U(z)\,.
  \end{split}
  \label{eq:LnU-n-geq-minus-one}
\end{align}
This is equivalent to 
\begin{equation} \label{eq:k_relabelling}
  \begin{aligned}
    k_{n \geq -1} =~ &+ \cL_n\\
                 &- \sum_{k=1}^n \binom{n+1}{k+1}z^{n-k}2k(\mu k-i\lambda)u_k\\
                 &+ \sum_{k=1}^n \binom{n+1}{k+1}z^{n-k}\sum_{m=1}^{k-1}m(k-m)u_m u_{k-m}\\
                 &- \sum_{k=0}^n \binom{n+1}{k+1} z^{n-k} \sum_{m=1}^\infty (m+k)u_{m+k}\partial_{u_m}.
  \end{aligned}
\end{equation}
Each line in this equation matches with the corresponding line in \eqref{eq:LnU-n-geq-minus-one} after renaming the summation variables $m \to m_1,k-m \to m_2$ in the third line and $m+k \to m,m-1 \to r$ in the last line. Equation \eqref{eq:k_relabelling} is then the expansion of
\begin{equation}
  k_{n\geq -1}
  =
  \mathcal L_n
  -(n+1)z^n(l_0-h)
  -
  \sum_{j=0}^{n-1}
  \binom{n+1}{j}
  z^j l_{-(n-j)}\,.
  \label{eq:kn-geq-minus-one-compact}
\end{equation}
For $n<-1$, we obtain through a similar relabelling
\begin{align}
  [L_n,U(z)]
  =
  {}&
  -\mathcal L_n U(z)
  \nonumber\\
  &-
  \sum_{m=1}^{\infty}
  (-1)^m
  \binom{-n+m-1}{m+1}
  z^{n-m}
  2m(\mu m-i\lambda)u_mU(z)
  \nonumber\\
  &+
  \sum_{m_1,m_2=1}^{\infty}
  (-1)^{m_1+m_2}
  \binom{-n+m_1+m_2-1}{m_1+m_2+1}
  z^{n-m_1-m_2}
  m_1m_2u_{m_1}u_{m_2}U(z)
  \nonumber\\
  &+
  \sum_{m=1}^{\infty}
  \sum_{r=0}^{m-1}
  (-1)^{m-r}
  \binom{-n+m-r-2}{m-r}
  z^{n+1-m+r}
  m u_m\partial_{u_{r+1}}U(z).
  \label{eq:LnU-n-less-minus-one}
\end{align}
This is the expansion of
\begin{equation}
  k_{n<-1}
  =
  \mathcal L_n
  -(n+1)z^n(l_0-h)
  +
  \sum_{k=1}^{\infty}
  (-1)^k
  \binom{-n-1+k}{k+1}
  z^{n-k}l_{-k}\,.
  \label{eq:kn-less-minus-one-compact}
\end{equation}
This concludes the derivation.

\section{Kač determinant in the oscillator formalism}
\label{sec:Kac-determinant}

We show in this section that the linear systems of equations determining the oscillator wavefunctions are solvable precisely for all non-degenerate Virasoro representations.

\subsection{Single-oscillator wavefunctions and the Gram matrix}
In the derivation of the four-point conformal block in the oscillator formalism, it is necessary to solve a linear system of equations \eqref{eq:coeff-F}, repeated here for convenience:
\begin{equation*}
  M \vec F_\psi = \vec B_\psi.
\end{equation*}
This system of equations can be solved for all coefficients if $\det(M) \neq 0$.
We now want to show that $\det(M) = 0$ only if the exchanged operator is degenerate.
In the textbook CFT treatment \cite{DiFrancesco:1997nk}, an operator is identified as degenerate if the determinant of the Gram matrix $G$ vanishes.
The matrix elements of $G$ are defined as
\begin{equation}
  G_{ij} = \bra{h}(L_{-i})^\dagger L_{-j}\ket{h}
  \label{eq:Gram-matrix}
\end{equation}
where $L_{-i},L_{-j}$ again denotes a string of Virasoro generators in shorthand notation.
In the following, we will show that $G$ is given as a product of $M$ and its complex conjugate such that $\det(G)=0$ is equivalent to $\det(M)=0$.
To do so, let us insert a resolution of the identity operator twice into \eqref{eq:Gram-matrix},
\begin{equation}
  \begin{aligned}
    G_{ij} &= \int[\mathrm dU][ \mathrm dV] \bra{h}(L_{-i})^\dagger\ket{\bar{v}}\bra{v}\ket{\bar{u}}\bra{u}L_{-j}\ket{h}\\
           &= \int[\mathrm dU][\mathrm dV] ((\bar{l}^{(v)}_{i})^\dagger 1)\braket{v}{\bar{u}} (l^{(u)}_{-j}1)\\
           &= \sum_k \frac{1}{(u_k,u_k)} \int[\mathrm dU][\mathrm dV] ((\bar{l}^{(v)}_{i})^\dagger 1)\bar{u}_kv_k (l^{(u)}_{-j}1).
  \end{aligned}
  \label{eq:Gram-matrix-rewritten}
\end{equation}
Here, in the second line we have used the representation of the Virasoro generators acting on an oscillator state and $\braket{h}{\bar{v}} = \braket{u}{h} = 1$.
In the last line, the inner product $\braket{v}{\bar{u}}$ has been expanded into monomials $u_k = u_{k_r}\dots u_{k_1}$, using that monomials form a complete orthogonal basis.
Since the monomials are not normalized, we have to add the normalization factor $\frac{1}{(u_k,u_k)}$ where the inner product is defined by $(f,g) = \int[\mathrm dU] \overline{f(U)}g(U)$.
Note also that $(\bar{l}^{(v)}_i)^\dagger = (\bar{l}_{i_p}^{(v)}\dots\bar{l}_{i_1}^{(v)})^\dagger = \bar{l}_{-i_1}^{(v)}\dots\bar{l}_{-i_p}^{(v)}.$
In order to relate this to the matrix $M$, we introduce the shorthand notation
\begin{equation}
  M_{ij} = l_{i_p}\dots l_{i_1} u_{j_q}\dots u_{j_1} := l_i u_j
\end{equation}
where $i,j$ are integer partitions of the level $n = \sum_k i_k = \sum_k j_k$ and $l_i,u_j$ denote a string of Virasoro generators and oscillators respectively.
We then have
\begin{equation}
  M_{ij} = l_i u_j = (1,l_i u_j) = ((l_{i})^\dagger 1,u_j)
\end{equation}
as follows from the definition of the inner product.
Using the definition of the inner product, we can therefore write
\begin{equation}
  \sum_k \frac{1}{(u_k,u_k)} M_{ik}\bar{M}_{kj} = \sum_k \frac{1}{(u_k,u_k)} \int[\mathrm dU][\mathrm dV] ((\bar{l}^{(v)}_{i})^\dagger 1)v_k\bar{u}_k (l^{(u)}_{-j}1).
\end{equation}
Comparing with \eqref{eq:Gram-matrix-rewritten}, we see that
\begin{equation}
  G = M N \bar{M}
  \label{eq:decomposition-Gram-matrix}
\end{equation}
where $N$ is the diagonal matrix of normalization factors for the monomials, $N_k = 1/(u_k,u_k)$.
This implies
\begin{equation}
  \det(G) = \frac{\det(M)\det(\bar{M})}{\prod_k (u_k,u_k)} = 0 \quad \Leftrightarrow \quad \det(M) = 0
\end{equation}
as claimed above.
Of course, the same arguments apply to the $\chi$ wavefunction.

\subsection{Wavefunctions with several oscillator variables}
The proof immediately generalizes to the case of a wavefunction with two oscillator variables.
In that case, the oscillator wavefunction obeys the differential equation \cite{Besken:2019bsu}
\begin{equation}
  (-kh_1-l_0^{(U)}+\bar{l}_0^{(\overline{V})} + l_k^{(U)} - \bar{l}_{-k}^{(\overline{V})})F_\Omega(U,\bar{V}) = 0.
\end{equation}
Expanding the oscillator wavefunction in terms of contributions from each level,
\begin{equation}
  F_\Omega(U,\bar{V}) = \sum_{n,m = 0}^\infty F_{n,m}(U,\bar{V}),
\end{equation}
where $F_{n,m} = \sum_{i,j} u_i \bar{v}_j$ with $\sum_k i_k = n,\sum_k j_k = m$, the equations at fixed level $(n,m)$ become
\begin{equation}
  l_k^{(U)}F_{n+k,m} - \bar{l}_{-k}^{(\overline{V})}F_{n,m-k} = (kh_1 + l_0^{(U)} - \bar{l}_0^{(\overline{V})})F_{n,m} = (kh_1+h_U-h_V+n-m)F_{n,m}
  \label{eq:two-oscillators-fixed-level}
\end{equation}
where by definition $F_{n,m} = 0$ for $n<0$ or $m<0$.
The structure of the equations which one needs to solve mirror those for a single oscillator.
For instance, for $m=0$ \eqref{eq:two-oscillators-fixed-level} becomes
\begin{equation}
  l_k^{(U)}F_{n,0} = (kh_1 + h_U-h_V+n-k)F_{n-k,0} := \gamma_{k,n-k,0}F_{n-k,0} \quad (k > 0),
\end{equation}
which yields by recursive application of Virasoro generators
\begin{equation}
  M \vec F_\Omega = \vec C_\Omega
  \label{eq:coeff-F-Omega}
\end{equation}
where $M$ is defined in \eqref{eq:definition-matrix-M}, $F$ is a vector of coefficients for the monomials of level $(n,0)$ in $F_{n0}$ and
\begin{equation}
  C_i = \gamma_{i_p,0,0}\gamma_{i_{p-1},i_p,0}\gamma_{i_{p-2},i_p+i_{p-1}}\dots\gamma_{i_2,n-i_1-i_2,0}\gamma_{i_1,n-i_1,0}.
\end{equation}
This equation obviously admits a solution whenever $\det(M) \neq 0$, giving the same conditions as above.
For $m>0$, \eqref{eq:two-oscillators-fixed-level} can be solved recursively by inserting the solutions for levels $(r,s)$ with $r<n$ or $s<m$.
This yields a similar system of linear equations as considered above,
\begin{equation}
    (M^{(U)} \otimes M^{(V)}) \vec F_\Omega = \vec B_\Omega
\end{equation}
where $\vec B_\Omega$ must be determined recursively from \eqref{eq:two-oscillators-fixed-level} and the matrix on the left-hand side is given by a tensor product with
\begin{equation}
  M^{(U)}_{ij} = l_i^{(U)} u_j, \quad M^{(V)}_{ij} = \bar{l}_i^{(V)} \bar{v}_j
\end{equation}
in the same shorthand notation as above.
Given that
\begin{equation}
  \det(M^{(U)} \otimes M^{(V)}) = \det(M^{(U)})^{p(m)} \det(M^{(V)})^{p(n)},
\end{equation}
the system of linear equations admits a solution whenever $\det(M^{(U)}),\det(M^{(V)}) \neq 0$, that is neither of the two internal operators is degenerate.

For a wavefunction with three oscillator variables, the equation that needs to be solved is given by
\begin{equation}
  \left\{
    \begin{aligned}
      \left[l_k^{(U)} - \bar{l}_{-k}^{(\overline{W})} - \sum_{l=0}^{k-1} \binom{k+1}{l}l_{-(k-l)}^{(V)} - l_0^{(U)} - kl_0^{(V)} + \bar{l}_0^{(\overline{W})}\right]F_\Xi &= 0, \quad (k \geq -1),\\
      \left[l_k^{(U)} - \bar{l}_{-k}^{(\overline{W})} + \sum_{l=1}^{\infty} (-1)^l \binom{-k-1+l}{l+1}l_{-l}^{(V)} - l_0^{(U)} - kl_0^{(V)} + \bar{l}_0^{(\overline{W})}\right]F_\Xi &= 0, \quad (k < -1).\\
    \end{aligned}
  \right.
  \label{eq:three-oscillators}
\end{equation}
This is obtained from
\begin{equation}
  (l_n^{(U)} - \bar{l}_{-n}^{(\overline{W})} + k_n^{(z,V)})\Xi = 0
\end{equation}
by rescaling $u_n \to u_n z^n, v_n \to v_n z^{-n}, w_n \to w_n z^{-n}$ and defining $\Xi = z^{h_U-h_V-h_W}F_\Xi$.
Expanding in contributions from fixed level, \eqref{eq:three-oscillators} becomes
\begin{equation}
  \left\{
    \begin{aligned}
      &l_k^{(U)}F_{n+k,m,p} - \bar{l}_{-k}^{(\overline{W})} F_{n,m,p-k} - \sum_{l=0}^{k-1}\binom{k+1}{l} l_{-(k-l)}^{(V)}F_{n,m-k+l,p}\\
      &\quad = (h_U + k h_V - h_W + n + km-p)F_{nmp}, \quad (k \geq -1),\\
      &l_k^{(U)}F_{n+k,m,p} - \bar{l}_{-k}^{(\overline{W})} F_{n,m,p-k} + \sum_{l=1}^{p}(-1)^l\binom{-k-1+l}{l+1} l_{-l}^{(V)}F_{n,m-l,p}\\
      &\quad = (h_U + k h_V - h_W + n + km-p)F_{nmp}, \quad (k < -1).\\
    \end{aligned}
  \right.
\end{equation}
This equation can also be solved recursively to obtain an equation of the form \eqref{eq:coeff-F} with
\begin{equation}
  M \to M^{(U)} \otimes M^{(V)} \otimes M^{(W)}.
\end{equation}
The fact that $M$ is given as a tensor product again implies that $\det(M) \neq 0$ if $\det(M^{(U,V,W)}) \neq 0$, giving the same condition that the internal operators have to be non-degenerate in order to find a solution for all coefficients.
The same applies to the $\Lambda$ wavefunction.

\subsection{Semiclassical limit}
Let us now consider the semiclassical limit.
For a wavefunction with a single oscillator variable, the system of equations to be solved at each level takes the form
\begin{equation}
    M_\text{cl.}\vec S = \vec B_\text{cl.}
    \label{eq:coeff-eq-semiclassical}
\end{equation}
where $S$ is a vector of coefficients of the monomial expansion of the exponent $S$ in the exponential ansatz \eqref{eq:exponential-ansatz-wavefunction} for the wavefunction while the matrix $M$ is obtained by nested application of the semiclassical oscillator operators defined in \eqref{eq:semiclassical-oscillator-operators}
\begin{equation}
    M_{ij,\text{cl.}} = \left.l_{i_p,\text{cl.}}[l_{i_{p-1},\text{cl.}}[\dots l_{i_1,\text{cl.}}[\rho_{j_q}\dots\rho_{j_1}]]]\right|_{\rho=0}.
\end{equation}
Note that the right hand side of \eqref{eq:coeff-eq-semiclassical} now includes coefficients of lower levels which are obtained recursively.
Uniqueness of the solution is established by an inductive argument if $\det(M_\text{cl.}) \neq 0$.
From \eqref{eq:coeff-eq-semiclassical}, it is straightforward to see that $M$ takes on triangular form such that
\begin{equation}
    \det(M_\text{cl.}) = \prod_j \prod_p (j_p + i\alpha)
\end{equation}
where $j$ is a partition of the level $k = \sum_p j_p$.
Thus, zeros of $M$ are located at $g = 1 + \alpha^2 = 1-r^2$ for positive integers $r$, showing that there is a unique solution of \eqref{eq:coeff-eq-semiclassical} for $h = \mu^2g >0$.
As expected, the location of the zeros corresponds to the appearance of a degenerate operator in the semiclassical limit with conformal weight $h = \mu^2(1-r^2) + O(\mu^0)$.
The same arguments as above apply to wavefunctions with multiple oscillator variables.

\section{Semiclassical vacuum block}
\label{sec:semiclassical-vacuum-block}
While in the main text, we have considered semiclassical conformal blocks where the weight of all internal and external operators scales with the central charge, for some applications semiclassical vacuum blocks are needed where one or multiple internal weights vanish.
In this section, we explain why the exponentiation proof does not straightforwardly extend to this setting based on the example of the four-point block on the sphere.

In order to show that the semiclassical vacuum block is well-defined, it is necessary to show that the semiclassical limit and vacuum limit both exist and commute,
\begin{equation}
    -\lim_{c \to \infty} \lim_{\gamma \to 0} \frac{6}{c}\log \left(\begin{tikzpicture}[scale=0.33,baseline={(0,0)}] \def\sizept{5pt}
        \draw (-2,1) node[left] {$h_1$} -- (-1,0);
        \draw (-2,-1) node[left] {$h_1$} -- (-1,0);
        \draw (-1,0) -- node[midway,above] {$\gamma c$} (1,0);
        \draw (1,0) -- (2,1) node[right] {$h_3$};
        \draw (1,0) -- (2,-1) node[right] {$h_3$};
         \fill (-2,1) circle (\sizept);
        \fill (-2,-1) circle (\sizept);
        \fill (2,1) circle (\sizept);
        \fill (2,-1) circle (\sizept);
    \end{tikzpicture}\right) = -\lim_{\gamma \to 0} \lim_{c \to \infty} \frac{6}{c}\log \left(\begin{tikzpicture}[scale=0.33,baseline={(0,0)}] \def\sizept{5pt}
        \draw (-2,1) node[left] {$h_1$} -- (-1,0);
        \draw (-2,-1) node[left] {$h_1$} -- (-1,0);
        \draw (-1,0) -- node[midway,above] {$\gamma c$} (1,0);
        \draw (1,0) -- (2,1) node[right] {$h_3$};
        \draw (1,0) -- (2,-1) node[right] {$h_3$};
        \fill (-2,1) circle (\sizept);
        \fill (-2,-1) circle (\sizept);
        \fill (2,1) circle (\sizept);
        \fill (2,-1) circle (\sizept);
    \end{tikzpicture}\right).
    \label{eq:semiclassical-vacuum-block}
\end{equation}
As we will show below, the proof given in the main text does not extend to this setting since the limits do not exist at the level of the oscillator wavefunctions.
There are nontrivial cancellations that occur after evaluating the integral over oscillator variables which conspire to give a well-defined semiclassical limit for the vacuum block even though the semiclassical vacuum wavefunctions are not well-defined.

\subsection{Vacuum block}
The existence of the semiclassical limit at $\gamma > 0$ has been established in the main text.
Let us now quickly establish that the limit $\gamma \to 0$ is well-defined at finite $c$ (so not in the semiclassical limit).
Consider the equation \eqref{eq:coeff-F} which determines the coefficients of the monomial expansion of the oscillator wavefunction at fixed level.
We need to show that this equation admits a unique solution for the coefficients $F_\psi$ in the limit $h \to 0$.
In the case $h_1 \neq h_2$, it is easy to see that already at level one the corresponding coefficient $F_{\{1\}}$ diverges in the limit $h \to 0$ as \eqref{eq:coeff-F} becomes
\begin{equation}
    (\mu + i\lambda)F_{\{1\}} = h + h_1 - h_2.
\end{equation}
While the LHS vanishes in the limit $\lambda \to i\mu$, the right-hand side is finite.
Therefore, the vacuum block is only well-defined if the external operators are pairwise equal, $h_1 = h_2$ and $h_3 = h_4$, as indicated in \eqref{eq:semiclassical-vacuum-block}.
In that case, for $i_p = 1$ we find that $\lim_{h \to 0} B_i = \lim_{h \to 0} \beta_{i_p,i_p} = 0$ (we have ordered $i_p \leq i_{p-1} \leq \dots \leq i_1$ w.l.o.g).
Moreover, for $i_p = 1$ we also find $\lim_{h \to 0} M_{ij} = \lim_{h \to 0} l_1u_1 = 0$.
Denote by $p(n)$ the number of partitions $n = \sum_k i_k$ of the integer $n$.
Since there are $p(n-1)$ cases of the $p(n)$ total ones where $i_p=1$ (decompose $n=1+n-1$ to see this), $p(n-1)$ equations within \eqref{eq:coeff-F} vanish on both sides in a naive $h \to 0$ limit.
We thus have to rescale these equations in order to find an equation which we can solve for the coefficients $F_\psi$.
Defining
\begin{equation}
    \tilde M_{ij} = \lim_{\lambda \to i \mu} \left\{\begin{aligned}
        &M_{ij}/(\lambda-i\mu), & i_p=1,\\
        &M_{ij}, & \text{otherwise},
    \end{aligned}\right.
    \qquad 
    \tilde B_i = \lim_{\lambda \to i \mu} \left\{\begin{aligned}
        &B_i/(\lambda-i\mu), & i_p=1,\\
        &B_i, & \text{otherwise},
    \end{aligned}\right.
\end{equation}
eq.~\eqref{eq:coeff-F} is equivalent to
\begin{equation}
    \tilde M \vec F_\psi = \vec{\tilde B}_\psi.
    \label{eq:coeff-F-vacuum}
\end{equation}
What remains is to prove that $\mathrm{det}(\tilde M) \neq 0$ in order for this equation to have a unique solution.
To see this, note that the determinant of the Gram matrix \eqref{eq:Gram-matrix} (the Kač determinant) is given by
\begin{equation}
    \det(G) \propto \prod_{r,s \geq 1, rs \leq n}(h-h_{r,s})^{p(n-rs)}
\end{equation}
where $h_{r,s}$ is the conformal weight of a degenerate operator with $h_{1,1} = 0$.
Thus, the $h=0$ zero of the Kač determinant occurs with multiplicity $p(n-1)$.
Due to \eqref{eq:decomposition-Gram-matrix}, this implies that the zero $\lambda = \pm i\mu$ also occurs with multiplicity $p(n-1)$ in $M$ and $\bar M$ respectively.
From $\mathrm{det}(M) = (\lambda - i\mu)^{p(n-1)} \mathrm{det}(\tilde M)$ due to the definition of $\tilde M$, it is then clear that $\mathrm{det}(\tilde M) \neq 0$.
This implies that the $h \to 0$ limit of the oscillator wavefunctions $\psi(z_1,z_2,U)$ and $\chi(z_1,z_2,\overline{U})$ with one oscillator variable are well-defined if $h_1=h_2$, implying also that the four-point block is well-defined.

\subsection{Non-existence of semiclassical vacuum wavefunction}
We now establish by explicit computation that the limits $c \to \infty$ and $\gamma \to 0$ of the oscillator wavefunction $\psi$ don't exist simultaneously.
The vacuum limit is given by
\begin{equation}
    \lim_{\gamma \to 0} \psi(x,h_1=h_2) = 1 + 2\mu u_1 x + (2 \mu u_2-2(h_1-2\mu^2)u_1^2)x^2 + O(x^3).
\end{equation}
Defining $u_m = \mu \rho_m$, $h_1 = \mu^2 g_1$, we find an ill-defined limit
\begin{equation}
    \lim_{\mu \to \infty} \lim_{\gamma \to 0} \frac{1}{\mu^2}\log(\psi(x,h_1=h_2)) = \lim_{\mu \to \infty} 2\rho_1 x + (2\rho_2 + 2 \mu^2(1-g_1)\rho_1^2) x^2 + O(x^3).
\end{equation}
Likewise, the reverse order of limits is also ill-defined.
While the semiclassical limit exists,
\begin{equation}
    \begin{aligned}
        &\lim_{\mu \to \infty} \frac{1}{\mu^2}\log(\psi(x,h_1=h_2)) = (1-i\alpha)\rho_1 x\\
        &\qquad + \left(\frac{4g_1-(1-i\alpha)(3+i\alpha)}{8(1+i\alpha)(2+i\alpha)}\rho_1^2 - \frac{4g_1+(1-i\alpha)(5+3i\alpha)}{4(2+i\alpha)}\rho_2\right)x^2 + O(x^3),
    \end{aligned}
\end{equation}
the vacuum limit $\alpha \to i$ of the semiclassical wavefunction does not exist.
However, the conformal block expanded to the same order in the conformal cross-ratio $x$ admits a well-defined semiclassical vacuum limit,
\begin{equation}
    \begin{aligned}
        &-\lim_{c \to \infty} \lim_{\gamma \to 0} \frac{6}{c}\log(\cF) = -\lim_{c \to \infty} \frac{(c-1)^2}{48c^2}g_1g_3 x^2 + O(x^3)\\
        &= -\lim_{\gamma \to 0}\lim_{c \to \infty} \frac{6}{c}\log(\cF) = -\lim_{\gamma \to 0} \frac{1}{8}\gamma x + \frac{(16 g_1g_3 + (4(g_1+g_3)+36)\gamma + 13\gamma^2)}{256(3+\gamma)}x^2 + O(x^3)\\
        &= -\frac{g_1g_3}{48}x^2 + O(x^3).
    \end{aligned}
\end{equation}
Therefore, the oscillator formalism is not well-suited to proving exponentiation of semiclassical vacuum blocks. 

In some cases, recursion relations for computing conformal blocks can be used to prove part of the question by showing that if both limits $c \to \infty$ and $\gamma \to 0$ exist, then they must commute \cite{Hartman:2013mia}.
However, the necessary recursion relations are available only for conformal blocks on the sphere \cite{Zamolodchikov1987,Artemev:2026cki} and the proof technique fails for other cases, such as the two-point block on the torus in the OPE channel.\footnote{The proof for the four-point block on the sphere from \cite{Hartman:2013mia} is based on decomposing the conformal block, which is a meromorphic function of the internal conformal weights, in a regular part times a sum over poles associated to degenerate representations of the Virasoro algebra. Using the residues of the poles which are known from the recursion relation one can show that the limits $c \to \infty$ and $\gamma \to 0$ commute in the singular part. Well-definedness of the semiclassical vacuum limit of the regular part can be read off from the recursion relation as well. However, for the two-point block on the torus the regular part does not admit a semiclassical vacuum limit as can be seen at third order in an expansion in the modular parameter $q$ and the cross-ratio $z_1-z_2$. Nevertheless, the semiclassical vacuum limit of the total conformal block $\lim_{c \to \infty} \lim_{\gamma \to 0} \frac{6}{c}\log$\begin{tikzpicture}[scale=0.25,baseline={(0,0)}] \def\sizept{4pt}
        \draw (-2,1) node[left] {$h_1$} -- (-1,0);
        \draw (-2,-1) node[left] {$h_1$} -- (-1,0);
        \draw (-1,0) -- node[midway,above] {$\gamma c$} (1,0);
        \draw (2,0) circle(1);
        \draw (3,0) node[right] {$h_q$};
        \fill (-2,1) circle (\sizept);
        \fill (-2,-1) circle (\sizept);
    \end{tikzpicture} is well-defined to the same expansion order.}
Thus, proving exponentiation of the semiclassical vacuum block remains an open question.

\bibliographystyle{JHEP}
\bibliography{bibliography.bib}

\providecommand{\href}[2]{#2}\begingroup\raggedright\begin{thebibliography}{10}

\bibitem{Zamolodchikov1986}
A.~Zamolodchikov, \emph{{Two-dimensional Conformal Symmetry and Critical
  Four-spin Correlation Functions in the Ashkin-Teller Model}}, {\emph{Zh.
  Eksp. Teor. Fiz.} {\bfseries 90} (1986) 1808}.

\bibitem{Zamolodchikov1987}
A.B.~Zamolodchikov, \emph{{Conformal symmetry in two-dimensional space:
  recursion representation of conformal block}}, {\emph{Teoreticheskaya i
  Matematicheskaya Fizika} {\bfseries 73} (1987) 103}.

\bibitem{Hartman:2013mia}
T.~Hartman, \emph{{Entanglement Entropy at Large Central Charge}},
  \href{https://arxiv.org/abs/1303.6955}{{\ttfamily 1303.6955}}.

\bibitem{Fitzpatrick:2014vua}
A.L.~Fitzpatrick, J.~Kaplan and M.T.~Walters, \emph{{Universality of
  Long-Distance AdS Physics from the CFT Bootstrap}},
  \href{https://doi.org/10.1007/JHEP08(2014)145}{\emph{JHEP} {\bfseries 08}
  (2014) 145} [\href{https://arxiv.org/abs/1403.6829}{{\ttfamily 1403.6829}}].

\bibitem{Fitzpatrick:2015zha}
A.L.~Fitzpatrick, J.~Kaplan and M.T.~Walters, \emph{{Virasoro Conformal Blocks
  and Thermality from Classical Background Fields}},
  \href{https://doi.org/10.1007/JHEP11(2015)200}{\emph{JHEP} {\bfseries 11}
  (2015) 200} [\href{https://arxiv.org/abs/1501.05315}{{\ttfamily
  1501.05315}}].

\bibitem{Besken:2019jyw}
M.~Be{\c{s}}ken, S.~Datta and P.~Kraus, \emph{{Semi-classical Virasoro blocks:
  proof of exponentiation}},
  \href{https://doi.org/10.1007/JHEP01(2020)109}{\emph{JHEP} {\bfseries 01}
  (2020) 109} [\href{https://arxiv.org/abs/1910.04169}{{\ttfamily
  1910.04169}}].

\bibitem{Desiraju:2024fmo}
H.~Desiraju, P.~Ghosal and A.~Prokhorov, \emph{{Proof of Zamolodchikov
  conjecture for semi-classical conformal blocks on the torus}},
  \href{https://arxiv.org/abs/2407.05839}{{\ttfamily 2407.05839}}.

\bibitem{Cho:2017oxl}
M.~Cho, S.~Collier and X.~Yin, \emph{{Recursive Representations of Arbitrary
  Virasoro Conformal Blocks}},
  \href{https://doi.org/10.1007/JHEP04(2019)018}{\emph{JHEP} {\bfseries 04}
  (2019) 018} [\href{https://arxiv.org/abs/1703.09805}{{\ttfamily
  1703.09805}}].

\bibitem{Pappadopulo:2012jk}
D.~Pappadopulo, S.~Rychkov, J.~Espin and R.~Rattazzi, \emph{{OPE Convergence in
  Conformal Field Theory}},
  \href{https://doi.org/10.1103/PhysRevD.86.105043}{\emph{Phys. Rev. D}
  {\bfseries 86} (2012) 105043}
  [\href{https://arxiv.org/abs/1208.6449}{{\ttfamily 1208.6449}}].

\bibitem{2020arXiv200303802G}
P.~{Ghosal}, G.~{Remy}, X.~{Sun} and Y.~{Sun}, \emph{{Probabilistic conformal
  blocks for Liouville CFT on the torus}},
  \href{https://doi.org/10.48550/arXiv.2003.03802}{\emph{arXiv e-prints} (2020)
  arXiv:2003.03802} [\href{https://arxiv.org/abs/2003.03802}{{\ttfamily
  2003.03802}}].

\bibitem{Arnaudo:2022ivo}
P.~Arnaudo, G.~Bonelli and A.~Tanzini, \emph{{On the Convergence of Nekrasov
  Functions}}, \href{https://doi.org/10.1007/s00023-023-01349-3}{\emph{Annales
  Henri Poincare} {\bfseries 25} (2024) 2389}
  [\href{https://arxiv.org/abs/2212.06741}{{\ttfamily 2212.06741}}].

\bibitem{LeFloch:2026xec}
B.~Le~Floch, \emph{{Convergence of Nekrasov instanton sum with adjoint
  matter}},  \href{https://arxiv.org/abs/2602.19425}{{\ttfamily 2602.19425}}.

\bibitem{Felder:2017rgg}
G.~Felder and M.~M{\"u}ller-Lennert, \emph{{Analyticity of Nekrasov Partition
  Functions}}, \href{https://doi.org/10.1007/s00220-018-3270-1}{\emph{Commun.
  Math. Phys.} {\bfseries 364} (2018) 683}
  [\href{https://arxiv.org/abs/1709.05232}{{\ttfamily 1709.05232}}].

\bibitem{Menotti:2025pqf}
P.~Menotti, \emph{{Convergence of classical conformal blocks}},
  \href{https://arxiv.org/abs/2512.18666}{{\ttfamily 2512.18666}}.

\bibitem{Antunes:2023kyz}
A.~Antunes, S.~Harris, A.~Kaviraj and V.~Schomerus, \emph{{Lining up a positive
  semi-definite six-point bootstrap}},
  \href{https://doi.org/10.1007/JHEP06(2024)058}{\emph{JHEP} {\bfseries 06}
  (2024) 058} [\href{https://arxiv.org/abs/2312.11660}{{\ttfamily
  2312.11660}}].

\bibitem{Buric:2021yak}
I.O.~Buri\'c, \emph{{Harmonic Analysis in Conformal and Superconformal Field
  Theory}}, Ph.D. thesis, University of Hamburg, Hamburg, 2021.
\newblock 10.3204/PUBDB-2021-04464.

\bibitem{Hadasz2010}
L.~Hadasz, Z.~Jaskolski and P.~Suchanek, \emph{{Recursive representation of the
  torus 1-point conformal block}},
  \href{https://doi.org/10.1007/JHEP01(2010)063}{\emph{JHEP} {\bfseries 01}
  (2010) 063} [\href{https://arxiv.org/abs/0911.2353}{{\ttfamily 0911.2353}}].

\bibitem{Kraus:2017ezw}
P.~Kraus, A.~Maloney, H.~Maxfield, G.S.~Ng and J.-q.~Wu, \emph{{Witten Diagrams
  for Torus Conformal Blocks}},
  \href{https://doi.org/10.1007/JHEP09(2017)149}{\emph{JHEP} {\bfseries 09}
  (2017) 149} [\href{https://arxiv.org/abs/1706.00047}{{\ttfamily
  1706.00047}}].

\bibitem{Alkalaev:2017bzx}
K.B.~Alkalaev and V.A.~Belavin, \emph{{Holographic duals of large-c torus
  conformal blocks}},
  \href{https://doi.org/10.1007/JHEP10(2017)140}{\emph{JHEP} {\bfseries 10}
  (2017) 140} [\href{https://arxiv.org/abs/1707.09311}{{\ttfamily
  1707.09311}}].

\bibitem{Rosenhaus:2018zqn}
V.~Rosenhaus, \emph{{Multipoint Conformal Blocks in the Comb Channel}},
  \href{https://doi.org/10.1007/JHEP02(2019)142}{\emph{JHEP} {\bfseries 02}
  (2019) 142} [\href{https://arxiv.org/abs/1810.03244}{{\ttfamily
  1810.03244}}].

\bibitem{Parikh:2019ygo}
S.~Parikh, \emph{{Holographic dual of the five-point conformal block}},
  \href{https://doi.org/10.1007/JHEP05(2019)051}{\emph{JHEP} {\bfseries 05}
  (2019) 051} [\href{https://arxiv.org/abs/1901.01267}{{\ttfamily
  1901.01267}}].

\bibitem{Fortin:2019dnq}
J.-F.~Fortin and W.~Skiba, \emph{{New methods for conformal correlation
  functions}}, \href{https://doi.org/10.1007/JHEP06(2020)028}{\emph{JHEP}
  {\bfseries 06} (2020) 028}
  [\href{https://arxiv.org/abs/1905.00434}{{\ttfamily 1905.00434}}].

\bibitem{Goncalves:2019znr}
V.~Gon\c{c}alves, R.~Pereira and X.~Zhou, \emph{{$20'$ Five-Point Function from
  $AdS_5\times S^5$ Supergravity}},
  \href{https://doi.org/10.1007/JHEP10(2019)247}{\emph{JHEP} {\bfseries 10}
  (2019) 247} [\href{https://arxiv.org/abs/1906.05305}{{\ttfamily
  1906.05305}}].

\bibitem{Parikh:2019dvm}
S.~Parikh, \emph{{A multipoint conformal block chain in $d$ dimensions}},
  \href{https://doi.org/10.1007/JHEP05(2020)120}{\emph{JHEP} {\bfseries 05}
  (2020) 120} [\href{https://arxiv.org/abs/1911.09190}{{\ttfamily
  1911.09190}}].

\bibitem{Fortin:2019zkm}
J.-F.~Fortin, W.~Ma and W.~Skiba, \emph{{Higher-Point Conformal Blocks in the
  Comb Channel}}, \href{https://doi.org/10.1007/JHEP07(2020)213}{\emph{JHEP}
  {\bfseries 07} (2020) 213}
  [\href{https://arxiv.org/abs/1911.11046}{{\ttfamily 1911.11046}}].

\bibitem{Alkalaev:2020yvq}
K.~Alkalaev and V.~Belavin, \emph{{More on Wilson toroidal networks and torus
  blocks}}, \href{https://doi.org/10.1007/JHEP11(2020)121}{\emph{JHEP}
  {\bfseries 11} (2020) 121}
  [\href{https://arxiv.org/abs/2007.10494}{{\ttfamily 2007.10494}}].

\bibitem{Hoback:2020pgj}
S.~Hoback and S.~Parikh, \emph{{Towards Feynman rules for conformal blocks}},
  \href{https://doi.org/10.1007/JHEP01(2021)005}{\emph{JHEP} {\bfseries 01}
  (2021) 005} [\href{https://arxiv.org/abs/2006.14736}{{\ttfamily
  2006.14736}}].

\bibitem{Fortin:2020zxw}
J.-F.~Fortin, W.-J.~Ma and W.~Skiba, \emph{{All Global One- and Two-Dimensional
  Higher-Point Conformal Blocks}},
  \href{https://arxiv.org/abs/2009.07674}{{\ttfamily 2009.07674}}.

\bibitem{Fortin:2020yjz}
J.-F.~Fortin, W.-J.~Ma and W.~Skiba, \emph{{Six-point conformal blocks in the
  snowflake channel}},
  \href{https://doi.org/10.1007/JHEP11(2020)147}{\emph{JHEP} {\bfseries 11}
  (2020) 147} [\href{https://arxiv.org/abs/2004.02824}{{\ttfamily
  2004.02824}}].

\bibitem{Fortin:2020bfq}
J.-F.~Fortin, W.-J.~Ma and W.~Skiba, \emph{{Seven-point conformal blocks in the
  extended snowflake channel and beyond}},
  \href{https://doi.org/10.1103/PhysRevD.102.125007}{\emph{Phys. Rev. D}
  {\bfseries 102} (2020) 125007}
  [\href{https://arxiv.org/abs/2006.13964}{{\ttfamily 2006.13964}}].

\bibitem{Haehl2020}
T.~Anous and F.M.~Haehl, \emph{{On the Virasoro six-point identity block and
  chaos}}, \href{https://doi.org/10.1007/JHEP08(2020)002}{\emph{JHEP}
  {\bfseries 08} (2020) 002}
  [\href{https://arxiv.org/abs/2005.06440}{{\ttfamily 2005.06440}}].

\bibitem{Alkalaev:2022kal}
K.~Alkalaev, S.~Mandrygin and M.~Pavlov, \emph{{Torus conformal blocks and
  Casimir equations in the necklace channel}},
  \href{https://doi.org/10.1007/JHEP10(2022)091}{\emph{JHEP} {\bfseries 10}
  (2022) 091} [\href{https://arxiv.org/abs/2205.05038}{{\ttfamily
  2205.05038}}].

\bibitem{Fortin:2023xqq}
J.-F.~Fortin, W.-J.~Ma, S.~Parikh, L.~Quintavalle and W.~Skiba, \emph{{One- and
  two-dimensional higher-point conformal blocks as free-particle wavefunctions
  in $ {\textrm{AdS}}_3^{\otimes m} $}},
  \href{https://doi.org/10.1007/JHEP01(2024)031}{\emph{JHEP} {\bfseries 01}
  (2024) 031} [\href{https://arxiv.org/abs/2310.08632}{{\ttfamily
  2310.08632}}].

\bibitem{Alkalaev:2023evp}
K.~Alkalaev and S.~Mandrygin, \emph{{Torus shadow formalism and exact global
  conformal blocks}},
  \href{https://doi.org/10.1007/JHEP11(2023)157}{\emph{JHEP} {\bfseries 11}
  (2023) 157} [\href{https://arxiv.org/abs/2307.12061}{{\ttfamily
  2307.12061}}].

\bibitem{Pavlov:2023asi}
M.~Pavlov, \emph{{Global torus blocks in the necklace channel}},
  \href{https://doi.org/10.1140/epjc/s10052-023-12166-7}{\emph{Eur. Phys. J. C}
  {\bfseries 83} (2023) 1026}
  [\href{https://arxiv.org/abs/2302.10153}{{\ttfamily 2302.10153}}].

\bibitem{Ammon:2024axd}
M.~Ammon, J.~Hollweck, T.~H{\"o}ssel and K.~W{\"o}lfl, \emph{{Conformal blocks
  in two and four dimensions from oscillator representations}},
  \href{https://doi.org/10.1007/JHEP05(2025)091}{\emph{JHEP} {\bfseries 05}
  (2025) 091} [\href{https://arxiv.org/abs/2406.19436}{{\ttfamily
  2406.19436}}].

\bibitem{Ammon:2025cdz}
M.~Ammon, J.~Hollweck, T.~H{\"o}ssel and K.~W{\"o}lfl, \emph{{Thermal $n$-Point
  Conformal Blocks in Four Dimensions from Oscillator Representations}},
  \href{https://arxiv.org/abs/2507.22974}{{\ttfamily 2507.22974}}.

\bibitem{Besken:2019bsu}
M.~Be{\c{s}}ken, S.~Datta and P.~Kraus, \emph{{Quantum thermalization and
  Virasoro symmetry}}, \href{https://doi.org/10.1088/1742-5468/ab900b}{\emph{J.
  Stat. Mech.} {\bfseries 2006} (2020) 063104}
  [\href{https://arxiv.org/abs/1907.06661}{{\ttfamily 1907.06661}}].

\bibitem{Ginsparg:1988ui}
P.H.~Ginsparg, \emph{{APPLIED CONFORMAL FIELD THEORY}},  in \emph{{Les Houches
  Summer School in Theoretical Physics: Fields, Strings, Critical Phenomena}},
  9, 1988 [\href{https://arxiv.org/abs/hep-th/9108028}{{\ttfamily
  hep-th/9108028}}].

\bibitem{Alkalaev:2015fbw}
K.B.~Alkalaev and V.A.~Belavin, \emph{{From global to heavy-light: 5-point
  conformal blocks}},
  \href{https://doi.org/10.1007/JHEP03(2016)184}{\emph{JHEP} {\bfseries 03}
  (2016) 184} [\href{https://arxiv.org/abs/1512.07627}{{\ttfamily
  1512.07627}}].

\bibitem{DiFrancesco:1997nk}
P.~Di~Francesco, P.~Mathieu and D.~Senechal, \emph{{Conformal Field Theory}},
  Graduate Texts in Contemporary Physics, Springer-Verlag, New York (1997),
  \href{https://doi.org/10.1007/978-1-4612-2256-9}{10.1007/978-1-4612-2256-9}.

\bibitem{Artemev:2026cki}
A.~Artemev and D.~Khromov, \emph{{WKB-asymptotics for multipoint Virasoro
  conformal blocks and applications}},
  \href{https://arxiv.org/abs/2603.08194}{{\ttfamily 2603.08194}}.

\end{thebibliography}\endgroup

\end{document}